\documentclass[12pt]{article}
\title{
QCD and Theory of Hadrons}
 \author{Yu.A.Simonov\\
State Research Center\\Institute of Theoretical and
Experimental Physics, \\
Moscow, Russia}
 \date{} \newcommand{\be}{\begin{equation}}
\newcommand{\ee}{\end{equation}}
\def\la{\mathrel{\mathpalette\fun <}}
\def\ga{\mathrel{\mathpalette\fun >}}
\def\fun#1#2{\lower3.6pt\vbox{\baselineskip0pt\lineskip.9pt
\ialign{$\mathsurround=0pt#1\hfil
##\hfil$\crcr#2\crcr\sim\crcr}}}

\newcommand{\ver}{\mbox{\boldmath${\rm r}$}}

\newcommand{\vep}{\mbox{\boldmath${\rm p}$}}

\newcommand{\vez}{\mbox{\boldmath${\rm z}$}}
\newcommand{\veS}{\mbox{\boldmath${\rm S}$}}
\newcommand{\veL}{\mbox{\boldmath${\rm L}$}}
\newcommand{\veR}{\mbox{\boldmath${\rm R}$}}
\newcommand{\ves}{\mbox{\boldmath${\rm s}$}}
\newcommand{\ven}{\mbox{\boldmath${\rm n}$}}
\newcommand{\verho}{\mbox{\boldmath${\rm \rho}$}}
\newcommand{\vexi}{\mbox{\boldmath${\rm \xi}$}}
\newcommand{\veta}{\mbox{\boldmath${\rm \eta}$}}
\newcommand{\veB}{\mbox{\boldmath${\rm B}$}}
\newcommand{\veE}{\mbox{\boldmath${\rm E}$}}
\newcommand{\veJ}{\mbox{\boldmath${\rm J}$}}

\begin{document}
\maketitle

\begin{abstract}
Nonperturbative QCD approach is systematically derived starting from
the QCD Lagrangian. Treating spin effects as a perturbation, one
obtains the universal effective Hamiltonian describing mesons,
hybrids and glueballs. \\
Constituent mass of quark and gluon is calculated via string tension.
The resulting spectrum of mesons, hybrids and glueballs obtained is
in good overall agreement with lattice data and experiment.
\end{abstract}

{\large \bf Contents}\\

\begin{enumerate}

\item Introduction.

\item  The QCD vacuum, confinement and chiral symmetry breaking.

\item   Perturbative and nonperturbative field configurations.
Asymptotic freedom and infrared
freezing.

\item Background perturbation theory and definition of hadron states.

\item  Relativistic path integrals and relativistic Hamiltonians.

\item Spectrum of light mesons, hybrids and glueballs. Regge trajectories.

\item Light--cone Hamiltonian and spectra for mesons.

\item Conclusions.

\end{enumerate}

\section{Introduction}

Hadron physics is a well--established field with enormous
experimental treasury of facts;
 on the theoretical side there are different models, some of these
 are well developed and have an impressive predicting power. The
 theoretical situation cannot be however considered satisfactory,
 taking into account that it is now clear that QCD is the real theory
 of strong interactions and can be used for selfconsistent
 construction of the hadron theory.  In addition to that new powerful
 methods in lattice simulations of QCD yield detailed information not
 only about spectrum of hadrons, but now also about scattering and
 hadronic matrix elements.

The reason for this  discrepancy was lack of analytic nonperturbative (NP)
methods in QCD, which could be used for systematic derivation of QCD-based
models. Recently the situation started to improve, and new powerful analytic
approaches have been suggested and checked versus experiment and lattice
data.  Remarkably many of the previous model results are derived now by
these approaches in a rather direct way from the first principles, i.e. from
the QCD Lagrangian, and it is gratifying that these features are supported by
experiment and lattice data. In addition new NP methods allow to
resolve old problems, which plauged existing  models. The
(incomplete) list of problems includes:

1) The problem of Regge slope. Relativistic quark models (RQM) predict for it
the value ($8\sigma)^{-1}$, where $\sigma$ is the string tension, while the
string models give $(2\pi\sigma)^{-1}$, which seems more realistic
from physical point of view.

2) The problem of Regge intercept. Both RQM and string models predict too
high values for the hadron masses, which is mended by introduction of large
negative constant in the Hamiltonian, of unknown dynamical origin.

3) Perhaps a more fundamental problem is the problem of constituent  mass of
quarks and gluons. The latter are introduced as a phenomenological input to
reproduce spectrum and static characteristics and are different in the
relativistic and nonrelativistic versions of the model. The dynamical origin
of constituent masses is unknown.

4)  The problem of quark radius and formfactors. In RQM ( to take a most
developed model) radius of hadrons comes out to be too small which is usually
mended by introduction of some effective quark radius, the origin
of this radius is not understood.

5) A very fundamental problem which lies at the basis of the whole QCD, is
the connection between the quark models of  hadrons (based on the minimal
number of degrees of freedom  -- those of valence quarks and gluons) and
high--energy scattering, including DIS, where bare (current) quark degrees of
freedom enter, including quark sea and gluon d.o.f. As an example, about one
half of energy--momentum is transported by a gluon cloud around
proton, while in the constituent quark model this gluon cloud is not
evident.

In this review we attempt to establish a basis for answering these questions,
using  new NP approaches in QCD. These include: 1) a new background field
theory, where NP fields play role of the background, and the
separation of perturbative and NP is made formally exact by  virtue
of the 'tHooft identity; 2) Relativistic path integral representation
for   quark and gluon Green's functions; 3) Method  of relativistic
Hamiltonians with auxiliary functions; 4) A very recent method of
effective contour--dependent quark Lagrangians, yielding a
simple picture of hadrons in the large $N_c$ limit.

The space limits of this review leave out many details of the formalism,
which can be found in the original literature. Moreover a very important part
of the hadron physics -- the strong hadron decays -- is left over completely
for future.

 \section{ The QCD vacuum, confinement and
 chiral symmetry breaking (CSB)}

QCD is believed to be the theory of strong interaction. The evidence
for that statement comes from comparison of  perturbative QCD and the
so-called QCD-motivated models with experiment and also from
lattice data.

There is some gap between models, treating nonperturbative (NP)
features of QCD, and QCD Lagrangian, which is now filled in by new
NP methods and Monte--Carlo calculations. It is the purpose of the
present review to describe these new NP methods, confronting their
results with experiment and lattice data, thus demonstrating that one
can indeed derive theory of hadrons from the first principles of QCD
with few assumptions, which can be checked independently.

To describe properties of the QCD vacuum and NP dynamics of quarks
and gluons one must consider the Euclidean QCD Lagrangian (since NP
configurations turn out to be Euclidean) and the QCD partition
function has the following form.
 \be
 Z=\int DAD\psi D\psi^+ {\rm exp} [L_0+L_1+L_{{\rm int}}]
 \ee
  where we are using Euclidean metric and define
  \be
  L_0=-\frac14\int d^4x(F^a_{\mu\nu})^2,
  \ee
  \be
  L_1=-i\int~^f\psi^+(x)(\hat \partial+m_f)~^f\psi(x)d^4x,
  \ee
  \be
  L_{\rm int}=\int~^f\psi^+(x) g\hat A(x)~^f\psi(x)d^4x.
  \ee

  Here and in what follows $~^f\psi_{a\alpha}$ denotes quark operator
  with flavour $f$, color $a$ and bispinor index $\alpha$.

The quantum theory of QCD is believed to be fixed by equations (1-4)
and one prescribed scale, e.g. $\Lambda_{QCD}$.

There are two main features of NP QCD, which make our world and
ourselves as we are:
 confinement and CSB.

 Confinement is a property of gluonic part of Lagrangian (2)  (for
 a review see [1]) and most easily defined for static quark $Q$ and
 antiquark $\bar Q$ neglecting light quarks, in which case the
 celebrated order parameter is the Wilson loop (to be often used
 below),
  \be \langle W(C)\rangle  = \int DA {\rm exp} L_0 \cdot P{\rm exp}  ig
 \int_C A_\mu dz_\mu.
  \ee
  We suppress here the gauge fixing and ghost terms for simplicity.
   For large contours $C$ one has Wilson criterium of confinement [1]
 \be
  \langle W(C)\rangle  = {\rm exp} (-\sigma S -\gamma P)
  \ee
   where $S$ is the
 minimal area inside contour $C$ and $\sigma$ is the string tension,
 while $P$ is the perimeter of the loop $C$. Note that theory must
 connect $\sigma$ and $\Lambda_{QCD}$ -- the only scale of QCD, how
 it is done on the lattice and in some models -- see in [2].

 Confinement in the form of area law (6) produces linear potential
 for heavy quark and antiquark at large distances, $V^{(C)}_{Q\bar
 Q}(R) =\sigma R$, and in section 5 we shall find  also other
 contributions to $V_{Q\bar Q}(R)$ due to perturbative gluon
 exchanges and spin--dependent parts. It is important that
 light--quark loop contributions are suppressed by powers of $1/N_c$
 and can be accounted for perturbatively.

 Whereas confinement enters directly via $W(C)$ in the heavy quark
 Green's function (see section 5), it is much more involved problem
 for light quarks, especially for  quarks in pions, since their
 Green's function cannot be expressed readily through $W(C)$.
 Physically this fact is connected to the Zitterbewegung  of light
 quark which is connected to the phenomenon of chiral symmetry
 breaking. To proceed one remarks that massless quarks can be split
 into left and right chiral parts, $\Psi(x)=\Psi_R+\Psi_L$  and
 $\Psi_R, \Psi_L$ enter Lagrangians (3) and (4) in the additive way,
 exemplifying the chiral symmetry of the group $U_L(N_f)\otimes
 U_R(N_f)$. It was shown in [3], that nonperturbative gluonic fields
 responsible for confinement, also produce CSB for light (massless)
 quarks. Concretely the appearance of the string (which transforms as
 Lorentz scalar) for a light quark is already a sign of spontaneous
 CSB, and the string appears as a solution of the nonlinear equation
 for the light--quark Green's function [3].

 For most mesons, however, not pions and kaons, chiral and spin
 effects can be treated as perturbation of the basic confinement
 dynamics. The corresponding Hamiltonian [4,5] looks much simpler,
 and reduces in some limiting cases to the familiar relativistic
 potential models [6,7,8]. We derive these Hamiltonians in section 5
 in the c.m. system.

 Of special interest in modern hadron spectroscopy are glueballs and
 hybrids. Here relativistic hamiltonians can be easily generalized to
 include gluon degrees of freedom, if gluon spin interaction is
 considered as perturbation. The spectrum obtained in this way is
 discussed in section 6 and is in good agreement with lattice data.
 One should note, that nowhere in our formalism we introduce
 constituent quark or gluon mass, and the only input mass parameters
 are string tension $\sigma$, and $\Lambda_{QCD}$ ( in addition to
 current quark masses $m_f$, renormalized at  the scale $1 GeV$).
 With such input one obtains a wide variety of hadron states and its
 properties in good overall agreement with experiment and lattice
 data. We also define and calculate the constituent mass of quark and
 gluon in terms of $\sigma$.

 \section{Perturbative and nonperturbative field configurations}

 One of basic mysteries of QCD is the fact, that gluon field plays
 two different roles:

 a) gluons are propagating, and at small distances this process can
 be described perturbatively, leading in particular  to color Coulomb
 interaction between quarks (antiquarks);

 b) gluons form a kind of condensate, which serves as a background
 for the propagating perturbative gluons and quarks. This background
 is Euclidean and ensures  phenomena  of  confinement and CSB.

 Correspondingly we shall separate  the total gluonic field $A_\mu$
 into perturbative part $a_\mu$ and nonperturbative (NP) background
 $B_\mu$:
 \be
 A_\mu=B_\mu+a_\mu
 \ee

 There are many questions about this separation, which may be
 answered now only partially, e.g. what exactly is the criterion of
 separation. Possible answer is that perturbative fields $a_\mu$ get
 their dimension from distance (momentum), and   therefore all
 correlators of   fields $a_\mu$ (in absence of $B_\mu$) are singular
 and made of
 inverse powers of ($x-y)$ and logarithms, where enters the only
 dimensional parameter of perturbative QCD -- $\Lambda_{QCD}$.
 Therefore evidently any dimensionful constant, like hadronic masses
 or string tension cannot be obtained as a perturbation series. In
 contrast to that, NP fields  $B_\mu$ have mass dimension due to the
 violation of scale invariance
 which is  intrinsically  present in the
 gluodynamics Lagrangian. The origin of separation (7) is clearly
 seen in the  solutions of nonlinear equations for field correlators
 [9]:  a perturbative  solution of those leads  to singular power-like
 field correlator, whereas at large distances there is a
 selfconsistent solution  of the equations, decaying exponentially
 with distance with arbitrary mass scale, since equations
 in  [9] are
 scale--invariant. Full solution including intermediate distances
 produces mixed perturbative--nonperturbative terms,  containing both
 inverse powers of distance and exponentials. For these terms
 criterion of separation fails.

 One can avoid formally the question of separation principle (and of
 double counting) using t'Hooft identity [10], which allows to
 integrate in (1) independently over $B_\mu$ and $a_\mu$:
 \be
 Z=\frac{1}{N'}\int DB_\mu \eta (B) D\psi D\bar \psi  D a_\mu
 e^{L(B+a)}
 \ee
 Here $L(A)=L_0+L_1+L_{int}$, and  $A_\mu$ is taken to be
 $B_\mu+a_\mu$. For the exact formalism starting from (8) we refer
 the reader to
 [11,12,2], and here we only quote the form of background
 propagator $G_{\mu\nu}$ of the gluon, which is found from $L(B+a)$
 and will be used in what follows,
 \be
 G_{\mu\nu}=(D^2_\lambda \delta_{\mu\nu} +2ig F_{\mu\nu})^{-1}
 \ee
 where $D_\lambda=\partial_\lambda-ig B_\lambda$, and $F_{\mu\nu} =
 \partial_\mu B_\nu-\partial_\nu B_\mu -ig [B_\mu, B_\nu]
 $. Neglecting NP background one calculates the perturbative static
 potential defined gauge--invariantly as
 \be
 V(r) =-\frac{1}{T} \lim_{T\to \infty} ln W(r,T)
 \ee
 Recent two--loop result in configuration space is [13]
 \be
 V(r) =-C_F\frac{\alpha_R(1/r)}{r},
 \ee
 and in momentum space one has [13]
 \be
 V(q) =-C_F\frac{4\pi \alpha_V(q)}{q^2}
 \ee
 with
 \be
 \alpha_V(q) =\alpha_{\bar M\bar S}(q)(1+a_1 \alpha_{\bar M\bar
 S}(q)+a_2\alpha^2_{\bar M\bar S})
 \ee
 and for the potential in position space one has [13,14]
 \be
 \alpha_R(1/r) =
 \alpha_{\bar M\bar S}(1/r)[1+c_1
 \alpha_{\bar M\bar S} (1/r) + c_2 \alpha^2_{\bar M\bar S})
 \ee
 where in quenched QCD $(n_f=0)$ one has
 \be
 c_1=1.832,~~ c_2=1.758
 \ee

 Employing the lattice measurement  of $\alpha (1/r)$
 in one obtains  [13,14]
 \be
 \Lambda^{(0)}_{\bar M\bar S}= 0.602/r_0,~~ r_0\approx 0.5 fm.
 \ee

 With this value of $\Lambda_R$ one obtains the renormalized
 $\alpha_R(1/r)$ which diverges at the Landau ghost pole, situated at
 $r\cong 0.2 fm$.

 Thus the perturbative potential is unacceptable in this form at
 $r\ga 0.2 fm$. However analysis of lattice data in [14,15] reveals,
 that Landau ghost pole is absent in the data.
    Potential models usually postulate saturating
 behaviour of $\alpha_R$ at large $r$ of the type [7]
 \be
 \alpha_R(1/r) = \left\{
 \begin{array}{ll}
 \alpha_R ({\rm one-loop)},& r<r_0\\
 {\rm const}=\alpha_R(r_0),~~ r\geq r_0
 \end{array}
 \right .
 \ee

 In this way one introduces a new parameter $
 \alpha_R(max)=\alpha_R(r_0)$, usually chosen around 0.4.

 Recently the physical mechanism producing freezing of $\alpha_R$ at
 large distances was identified in
 [11,12]. It is connected to the fact,
 that in the  confining background gluons cannot propagate very far
 from the sources, and typically this distance is invesely
 proportional to the corresponding hybrid excitation energy
 (which is of the order of $1.0 \div 1.2 GeV$).

 At large $N_c$ and in the Euclidean region, $q^2>0$,  one can predict
 this saturation (freezing) of  $\alpha_s$ theoretically to one loop
 as [12]
 \be
 \alpha(q) =\frac{4\pi}{\psi(\frac{q^2+M^2_0}{m^2})+ln
 \frac{m^2}{\Lambda^2}}
 \ee
 where $\psi$  is the Euler function, $\psi(z)=\Gamma'(z)/\Gamma(z),
 m^2$ is the scale of radial excited states at large $n$
 \be
 M^2_n= nm^2+M^2_0
 \ee
 and $M^2_0$ the lowest mass state in the channel.

 When the argument of  $\psi$ function is large, one has asymptotic
 representation
 \be
 \psi(z) =ln z -\frac{1}{2z} -\sum_{k=1}^\infty
 \frac{B_{2k}}{2kz^{2k}}
 \ee
 Correspondingly $\alpha(q)$ asymptotically assumes the form
 \be
 \alpha^{(as)} (q) =\frac{4\pi}{b_0ln \frac{q^2+M^2_0}{\Lambda^2}},~~
 b_0=\frac{11}{3} N_c-\frac{2}{3} n_f
 \ee
 A naive  "explanation" of (21) is that gluon acquires the mass
 $M_0=m_g$ which eliminates Landau ghost pole.

 In reality gluon does not  have the mass (this would violate gauge
 invariance and render the theory nonrenormalizable), but due to
 confinement is connected by the string to the quarks and this
 creates the mass of the whole system -- the hybrid.

The two--loop asymptotic form of $\alpha$ in the position space looks
like [2,12]
\be
\alpha_R(1/r)=\frac{4\pi}{b_0\ln a(r)}\{1+\frac{b_1}{b_0}\frac{\ln\ln
a(r)}{\ln a(r)}\}^{-1}
\ee
where $a(r)\equiv \frac{M^2_0+r^{-2}}{\Lambda^2_R}$, and $b_1=102$.

Note that the combination $M_0^2+q^2, M_0^2+1/r^2$ in $a$ is the
renormalization group (RG), invariant, since both external momenta and
background fields ($gB_\mu)$,  defining $M_0$ are RG invariant [2,12]
(while $g$ and $B_\mu$ separately are not).

There are many phenomenological arguments in favour of freezing of
$\alpha_s$ at large distances, for a rewiew see [16] and more recent
discussion in [12]. A different approach to the freezing behaviour,
based on some assumed analytic properties of $\alpha$ has been
suggested in [17].

Summarizing phenomenological situation one can state that all known
facts are supporting  saturating $\alpha$ in Eucledian region at the
level $\alpha(max) \leq 0.8$.

Recent lattice data allow a more direct computation of $\alpha(1/r)$
and  thereby the freezing phenomenon. In [18] the form (22) was used
for $\alpha_F$, entering the force between static quarks (which is
easier to measure on the lattice)
\be
F(r) =\frac{4}{3} \frac{\alpha_F(r)}{r^2}
\ee
The lattice measurements of [15] are compared in Fig.1
 of [18] to the
theoretical curve (23) with $M_0=1 GeV$ and $\Lambda_R=280 MeV$. One
can see there a good agreement, whereas the nonfreezing
behaviour $(M_0=0)$ clearly contradicts data at $r\geq 0.4 fm$.

As one can see from [18], lattice calculations give a clear direct
evidence for the freezing of the coupling constant, with the maximal
value
\be
\alpha^{Lattice}(max)\leq  0.5
\ee
A recent experimental evidence for freezing of $\alpha_s$ was found
in the detailed analysis of charmonium spectra in [19].

At this point it should be noted that freezing (saturation) is a
property of Euclidean (space--like) interactions. In the timelike
region $(q^2<0)$ Equation
(12) yields complex logarithm, which is
drastically important for time--like formfactors and Drell--Yan
processes. From physical point of view logarithmic branch points of
$\alpha_s$ are not relevant, and the analytic structure of $\alpha_s$
is better displayed by the Equation (18) valid in the large $N_c$
limit. In this limit all singularities of physical amplitudes are
poles, and in the physical scheme of $\alpha_s$ renormalization,
 when $\alpha_s$ is directly connected to some
physical amplitude, $\alpha_s$ should have only poles, which are
clearly displayed in (18). At large $|q^2|, q^2\gg m^2$, one can
average over many poles and return to the  complex logarithmic
dependence in (12), in approximate agreement with standard $\alpha_s$
expression, where $M_0^2 \equiv 0.$

So far we explored purely perturbative expressions for static
potential or modified due to the NP background. Now we turn to the
purely NP static interaction, i.e. to the confining potential. To
this end one rewrites  the NP background part
of the Wilson loop using the nonabelian Stokes theorem [20] as
$$
\langle W(B)\rangle  =\frac{1}{N_c} tr \langle P~ exp~ig \int_C B_\mu dx_\mu\rangle _B=
 $$
 \be
 =\frac{1}{N_C}tr\langle  exp~ig \int_S d\sigma_{\mu\nu} F_{\mu\nu} (u,
 x_0)\rangle _B.
 \ee

 Here $S$ is a surface bounded by contour $C$ and $x_0$ -- an
 arbitrary point on $S$, we also defined:
 \be
 F_{\mu\nu}(u,x_0)= \phi(x_0,u) F_{\mu\nu}(u) \phi(u,x_0)
 \ee
 and $\Phi(x,y)$ is the parallel transporter along some contour from
 $y$ to $x$.
 \be
 \phi(x,y) = P exp~ig \int^x_y B_\mu(z) dz_{\mu}
 \ee

 To evaluate the last integral in (25) one can  use  the cluster
 expansion theorem [21]
 \be
 \langle W(B)\rangle  =\frac{1}{N_c} tr exp \sum^{\infty}_{n=1}\frac{(ig)^n}{n!}
 \int_S d\sigma(1)...\int_S d\sigma(n)\ll F(1)... F(n)\gg
 \ee
 Here $F(k)\equiv F_{\mu_k\nu_k}(u^{(k)},x_0), d\sigma(k)\equiv
 d\sigma_{\mu_k\nu_k}(u^{(k)})$.

 The double brackets in (28) denote the cumulant, or connected
 average, e.g. for $n=2$ \be \ll F(1)F(2)\gg =\langle F(1)F(2)\rangle
 -\langle F(1)\rangle \langle F(2)\rangle  \ee

 For more details and definitions of cluster expansion see [21].

Throughout this review we shall consider cumulants or field
correlators (FC) $
 \langle\langle F(1)...F(n)\rangle\rangle$
 as the basic NP input, which defines all dynamics of the
 quark--gluon systems, both for relativistic or nonrelativistic
 situations and potential or nonpotential regimes. It is a rigorous
 and explicit language which allows to derive properties of
 confinement and CSB from the structure of cumulants, as will be
 shown below. On the other hand, FC are gauge--invariant and
 Lorentz--covariant quantities which can be found independently, both
 analytically and on the lattice. Analytic studies of the lowest
 cumulant $\langle FF\rangle $ have been done recently in [9],  using selfcoupled
 equations which exist for large $N_c$.   Lattice studies are by now
 numerous for $\langle FF\rangle $ [22-25] and also quartic FC--cumulant
 $
 \ll FFFF\gg$ --
 was studied in [26].

 Let us discuss first confinement, i.e. the area law (6) in terms of
 FC. To this end we rewrite (28) as
 \be
  \langle W(B)\rangle =\frac{1}{N_c} tr exp
 [-\frac12\int_Sd\sigma_{\mu\nu}(u)\int_sd\sigma_{\rho\sigma}(u)
 \Lambda_{\mu\nu,\rho\sigma}(u,v,C)]
 \ee
 where the global correlator $\Lambda$ is introduced,
 $$
 \Lambda_{\mu\nu,\rho\sigma} (u,v,C) \equiv
 g^2\ll F_{\mu\nu}(u,x_0) F_{\rho\sigma} (v, x_0)\gg
 -
 $$
 \be
 -2 \sum^\infty_{n=3}\frac{(ig)^n}{n!} \int d\sigma (3)... d\sigma
 (n) perm\ll F(u) F(v) F(3)... F(n)\gg
 \ee
 and $perm$ denotes sum over permutations of  $F(u), F(v)$ and other
 terms in the cumulant.

 Consider now the lowest term in (31), $\ll FF\gg$.

 Both from lattice measurements [
 22-25] and analytic  study [9] this term
 decays exponentially  with  $|u-v|$ and the characteristic
 correlation length $T_g$, which is called the gluon correlation
 length, is rather small,

 \be
 T_g=0.2\div 0.3 fm
 \ee
 This quantity is a basic characteristic of the QCD vacuum, defining
 different dynamical regimes for systems of the range $R$, in cases
 $R\langle T_g$ or $R\gg T_g$, as will be shown in chapter 5. For  an
 earlier discussion of $T_g$ see [27].

 Consider now Wilson loop of large radius $R, R\gg T_g$. Then in the
 integral (30) the generic situation is when $|u-x_0|, |v-x_0|$ of
 the order of $R$. In this case the correlator $\langle
 F_{\mu\nu}(u,x_0) F_{\rho\sigma}(v, x_0)\rangle$ does not depend on
 $x_0$ (up to the terms $O(T_g^2/R^2)$) and can be rewritten as the
 function of $|u-v|$. The general decomposition then can written as
 [28]
   \be
   g^2tr\langle F_{\mu\nu}(x_1)\Phi(x_1x_2)F_{\lambda\sigma}
   (x_2)\Phi(x_2,x_1)\rangle=N_c[(\delta_{\mu\lambda}\delta_{\nu\sigma}-
   \delta_{\mu\sigma}\delta_{\nu\lambda})D(u)+
   \ee
   $$
   +\frac{1}{2}(\frac{\partial}{\partial
    x_{1\mu}}u_{\lambda}\cdot\delta_{\nu\sigma}
   +\frac{\partial}{\partial
    x_{1\lambda}}u_{\mu}\delta_{\nu\sigma}+perm) D_1(u)]
   $$
 Insertion of (33) in the integral
  yields readily the Wilson area law
  \be
  \langle W(B)\rangle= exp (-\sigma S-\gamma P)
  \ee
  where $\sigma $ is expressed through the function $D(x)$, while the
  perimeter term $\gamma P$ is due to both $D$ and $D_1$,
  \be
  \sigma =\frac{1}{2} \int d^2 x D(x)+...
  \ee
  The ellipsis in (35) implies possible contribution of higher
  correlators in (31). For all of them, as well as for the Gaussian
  correlator, a condition necessary to contribute to the string
  tension $\sigma$ is that they should contain "Kronecker component",
  like $D(x)$, which is a coefficient of product of Kronecker
  $\delta$ symbols.

  As a consequence this term violates Abelian Bianchi identities [28]
  (which can be checked acting on both sides of (33) with operators
  $\varepsilon_{\alpha\beta\mu\nu}\frac{\partial}{\partial x_\beta}$)
  and hence is connected with the contribution of abelian projected
  monopoles. This is one of the possible illustration of the physical
  mechanism of confinement induced by correlator $D(x)$ (and similar
  higher correlators).

  Let us come back to the gluonic correlation length $T_g$, which
  together with $\sigma$ constitute the basic nonperturbative scales.
  Actually one can calculate $T_g$ through $\sigma$, as it was done
  in [9] using equations for correlators, and through gluon
  condensate  using QCD sum rules [29].  In both cases
  $T_g\ll \Lambda_{QCD}$ and $T_g\ll\frac{1}{\sqrt{\sigma}}$, which
  means that $T_g$ signifies a new  physical scale. The physical
  meaning of $T_g$ can be understood, when one calculates the field
  distribution in the QCD string [26], where Gaussian correlator
  $D(x)$ yields a very good description of lattice data and reveals
  that $T_g$ defines the width $l$ of the QCD string (where fields
  decrease by 50\%), namely $l\sim 2T_g$. From lattice data [26]
  one has
  \be
  l=0.4\div 0.5 fm,~~T_g=0.2\div
  0.3fm.
  \ee
   In what follows we shall sometimes use the limit $T_g\to
  0$, which we shall call "the string limit". Actually this is the
  limit of thin strings which is as a rule much simpler than dynamics
   of the realistic QCD string of finite width ( the same limit is
  always assumed in the theory of strings and superstrings).

  The finite value of $T_g$ brings about specific effects. E.g. for
  the potential of static quarks $V(r)$, which is obtained from the
  Wilson loop as in (10), one can express $V(r)$ through $D(x)$  and
  higher correlators,  and analyze the large and small--$r$
  behaviour, which yields [28]
  \be
  V(r) = \sigma r
  -c_0, ~~r\to \infty;~~~ V(r) \sim c_2 r^2+c_4r^4+...,~~ r\la
  T_g
  \ee
  where $
  c_0\sim \sigma T_g,~~c_2$ is expressed through $D(x)$, $c_4$
  through quartic correlator and so on. Eq.(37) demonstrates that NP
  fields ($B_\mu)$ are soft and yield analytic behaviour of $V(r)$ at
  $r\ll T_g$, corresponding to the OPE. Interference of perturbative
  and NP fields may change this picture at small $r$ yielding linear
  term in $r$, as was shown in [30]. For more discussion in
  connection with OPE and QCD sum rules see [31]. We shall not go
  into details of this phenomenon, but  mention that linear behaviour
  at small $r$ is phenomenologically necessary for good description
  of fine structure of heavy quarkonia [19,32] and was found recently
  on the lattice [14].

  We end this section with the discussion of the validity of Gaussian
  approximation, when only quadratic correlator $\langle FF\rangle$
  is taken into account. There are arguments in favour of Gaussian
  approximation both from comparison with lattice data and from the
  structure of the method.

  The first arguments come from the analysis of field distribution
  inside the string made in [26], as was discussed above. There the
  largest omitted contribution (from the quartic correlator) amounts
  to less than few percent. Another comparison is for the static
  potentials in higher representations, which demonstrate in numerous
  lattice data [33]  (for earlier data see in [1]) a clear Casimir
  scaling.  This is in agreement with the Gaussian approximation,
  while the quartic term would violate the Casimir scaling [1],[34].
  From the numerical accuracy of the latter one can again deduce,
  that Gaussian approximation is valid within 10\% of accuracy, or
  even better.

  Finally one can estimate the parameter of cluster expansion as
  follows.
 All observables
are expressed in terms of connected FC of gluonic fields,
$\langle\langle F(1)...F(n)\rangle\rangle $. E.g.  the string tension
for heavy (static) quarks is an  infinite sum of FC of the field
$F_{14}\equiv E_1$ integrated over the plane $14$:  \be \sigma \sim
\sum_n\frac{g^n}{n!}\prod^{n-1} d^2r_i\langle\langle  E_1(0)
E_1(r_1)E_1(r_1+r_2)...E_1(\sum r)\rangle\rangle
\ee
One can identify parameter of expansion in the sum (38) to be (only
even  powers of $n$ enter the sum)
\be
\zeta = (\bar E_1 T^2_g)^2
\ee
where $T_g$ is the gluonic correlation length in the vacuum
defined by the exponential decay of FC, and
$\bar E^2_1 \cong g^2\langle (E^a_1)^2\rangle
= \frac{4\pi^2}{12} G_2 \approx (0.2
GeV)^2
$ while  $G_2$ is the standard gluon condensate.

Lattice calculations confirm that $T_g$ is rather  small [22-25],
indeed $T_g\approx 0.2 \div 0.3 fm$ and therefore  $\zeta$
is a good expansion parameter
\be
\zeta=0.04 \div 0.1
\ee

The regime (40) $\zeta\ll 1  $
 which seems to be characteristic of real
QCD, can be called the regime of the \underline{weak~~confinement}.
In this case the dynamics of quarks and gluons is adequately
described in all known cases by the lowest (Gaussian) correlator.

One should mention the negative feature of Gaussian approximation.
Whereas the global correlator $\Lambda_{\mu\nu, \rho\sigma}$ --
containing the infinite sum over reduced correlators (cumulants)
yields the area law for the Wilson law, not depending on the shape of
the surface being integrated in (31), and hence reducing to the
minimal surface,retaining only Gaussian correlator one gets as a
price for simplification the parasytic dependence on the shape of the
surface $S$ in (30). Therefore saying about the dominant role of the
Gaussian correlator and smallness of discarded terms one should
specify that estimates refer to the minimal surface. On the other
hand for an arbitrary surface of any weird shape the contribution of
higher correlators can be large and this is what exactly needed for
compensation of the extra area contribution of the Gaussian
correlator.

The situation here is the same as
in the QCD perturbation series, which depends on the normalization
mass $\mu$ for any finite number of terms of the series. This
unphysical dependence is usually treated by fixing $\mu$ at some
physically reasonable value $\mu_0$ of the order of the inverse size
of the system.

In what follows we shall {\bf not} use Gaussian approximation, expert
for calculation of spindependent  contributions, which amounts to a
relatively small correction in good agreement with  lattice data and
experiment.

\section{Background perturbation theory and definition of hadron
states}

To define the perturbation theory  series in $ga_\mu$ one starts from
(8) and  rewrites the Lagrangian as follows:
$$
L_{tot}=L_{gf}+L_{gh}+L(B+a)=
$$
\be
L_0+L_1+L_2+L_{int}+L_{gf}+L_{gh}
\ee
where $L_i$ have the form:
\begin{eqnarray}
L_2(a)&=&\frac{1}{2} a_{\nu}(\hat{D}^2_{\lambda}\delta_{\mu\nu} -
\hat{D}_{\mu}\hat{D}_{\nu} + ig \hat{F}_{\mu\nu}) a_{\mu}= \nonumber \\
&=&\frac{1}{2} a^c_{\nu}[D_{\lambda}^{ca}D_{\lambda}^{ad} \delta_{\mu\nu}
- D_{\mu}^{ca}D_{\nu}^{ad} - g~f^{cad}F^a_{\mu\nu}]a^d_{\mu}~~,
\end{eqnarray}
\begin{eqnarray}
D_{\lambda}^{ca} &=&
 \partial_{\lambda}\cdot \delta_{ca}+ g~f^{cba} B^b_{\lambda}
\equiv \hat{D}_{\lambda}
\nonumber
\\
L_0 &=& -\frac{1}{4}
 (F^a_{\mu \nu}(B))^2 ~;~~~ L_1=a^c_{\nu} D_{\mu}^{ca}(B) F^a_{\mu\nu}
\nonumber
\\
L_{int} &=& -\frac{1}{2} (D_{\mu}(B)a_{\nu} -D_{\nu}(B)a_{\mu})^a
g~f^{abc} a_{\mu}^b a_{\nu}^c - \frac{1}{4} g^2 f^{abc} a_{\mu}^b a_{\nu}^c
f^{aef} a^e_{\mu} a^f_{\nu}
\end{eqnarray}
\begin{eqnarray}
L_{ghost}=-\theta^+_a (D_{\mu}(B)D_{\mu} (B+a))_{ab} \theta_b.
\end{eqnarray}
To calculate Green's functions of any hadrons one can use  (8) and
write
$$
G_h(X;Y)= const \int DB_\mu\eta(B)D\psi D\bar \psi Da_\mu
\Psi^+_f(X)\Psi_{in} (Y) e^{L_{tot}}
$$
\be
\equiv \langle \Psi^+_f(X)\Psi_{in} (Y)\rangle _{B,\psi,a}
\ee
Here $\Psi_{f, in} $ are initial and final hadron states made of
$B,\psi,\bar\psi,a$. To calculate integral $D{a_\mu}$ one can use
perturbative expansion in $ga_\mu$, i.e. neglect in first
approximation $L_{int}$, containing terms  $a^3, a^4$ and take into
account only quadratic terms $L_2, L_{gf}, L_{gh}$.
contribution of $L_1$ was studied in [11] and it was shown there that
for the most processes (e.g. for hadron Green's functions) it can be
neglected (with accuracy of about better than 10\%).

It is convenient to choose background gauge for $a_\mu$, $D_\mu
a_\mu=0$, and take gauge transformation in the form

\be
B_\mu\to U^+(B_\mu+ \frac{i}{g} \partial_\mu)U, a_\mu\to U^+ a_\mu U
\ee
Now one can choose $\Psi_f, \Psi_{in}$ as gauge-invariant
 forms built from $B_\mu, a_\mu,\psi,\bar \psi$ and
 to average over $B_\mu,\psi,\bar \psi, a_\mu$ as shown in (45).
Integrating over $DB_\mu$ might seem an impossible adventure due to
unknown $\eta(B)$, but we shall see that it can be always
written as products of  Wilson loops with some insertions, which can
be treated in two ways:

i) or in  the form of the area law. Then contribution of fields
$B_\mu$ is  given  and fixed by the string tension $\sigma(B)$.
Equating $\sigma(B)= \sigma_{exper}$ one fixes contribution of the
background fields.

ii) Using cluster expansion, and expressing result through lowest
correlators, e.g. $\langle FF\rangle $, or $D(x)$ and $D_1(x)$. In this case
background is fixed by these functions taken as an input, e.g. from
lattice data.

The rest integrals, over $Da_{\mu} D\psi D\bar \psi$ are Gaussian and
can be
simply done. Now we turn to the construction of $\Psi_{f, in}$.

One can choose local or nonlocal forms for $\Psi_{in,f}$, the latter
are obtained by insertion of $\phi(x,y)$ (27) in the local
expressions, which are

$\Psi_{in,f}=\bar \psi \Gamma_i \psi$ for mesons, $\Gamma_i=1,
\gamma_5, \gamma_\mu, \gamma_\mu\gamma_5, \sigma_{\mu\nu}$

$\Psi_{in,f}=\bar \psi \Gamma_i f (a, Da)\psi $ for hybrids,
where $f$ is a polynomial, the simplest form is
$\bar \psi \gamma_i a_\mu\psi$

$\Psi_{in,f}=tr a\Lambda_2 a$ and $tr(\Lambda_3 a^3)$ for
glueballs made of 2 and 3 gluons  respectively. $\Lambda_2,
\Lambda_3$ are polynomials made of $D_\mu(B)$ and ensure the
needed tensor structure for given quantum numbers. In the
simplest case $\Lambda_2=\Lambda_3=1$.

$\Psi_{in,f}=e_{abc} \psi^a_{f_1\alpha} \psi^b_{f_2\beta}
 \psi^c_{f_3\gamma} K^{
 f_1\alpha,f_2\beta,f_3\gamma}$ for baryons, where $a,f_i,\alpha$ are
 color, flavour and Lorentz index respectively.

 It is clear that higher hybrid states for mesons and baryons are
 obtained by adding additional factors of $a_\mu$ in $\Psi_{in, f}$.

 Several comments are in order. Our definition of hybrids differs
 from that of the flux tube model [35], where hybrid state is a result
 of the string excitation. In the latter case the quantum numbers of
 the hybrid are fixed by the mode of the string. In our case gluon
 $a_\mu$ sitting on the string has its own spin and makes account for
 larger set of values\footnote{the author is grateful to
 Yu.Kalashnikova for a discussion of this point}.  Also dynamics is
 different and local state $\Psi_{in,f}$ for  hybrid in the flux tube
 model does not exist. The same differences exist for glueballs.
 In contrast to other definitions
  our definition of hybrids and glueballs is unique and is given
 in the field theoretical terms. The correspondence with the lattice
 is also clear: to the gluonic excitation in $\Psi_{in, f}$ there
 corresponds a plaquette on the parallel transporter, which is field
 strength $F_{\mu\nu}$ in the continuum limit, (or a shift
 corresponding to an extra link, which is $A_\mu$ or $D_\mu$ in the
 continuum).

 Insertion of given $\Psi_{in, f}$ in (45) yields after integration
 over $Da_\mu D\psi D\bar\psi$ the hadron Green's function, which can
 be written [36] respectively  for mesons
 $$
 G_M(X,Y)=\langle tr(\Gamma_i^{(f)}(X)G_q(X,Y)
 \Gamma_k^{(in)}(Y)G_{\bar q}(X,Y))\rangle _B
 $$
   \be
 -\langle tr(\Gamma_i^{(f)}(X)G_q(X,X))
 tr (\Gamma_k^{(in)}(Y)G_{\bar q}(Y,Y))\rangle _B,
   \ee
   where $G_{q,\bar q}$ is the quark (antiquark) Green's function,
   \be
   G_q(X,Y)=(-i\hat \partial-im-g\hat B)^{-1}_{X,Y},
   \ee
   for hybrids
   \be
 G_{hyb}(X,Y)=\langle tr(\Gamma_i^{(f)}(X)G_q(X,Y)
 G_g(X,Y)\Gamma_k^{(in)}G_{\bar q}(X,Y))\rangle _B,
 \ee
   where $G_g$ is the gluon Green's function(9).

   For glueballs one has
   \be
 G_{glueball}=\langle tr(\Lambda_2^{(f)}(X)G_g(X,Y)
 \Lambda_2^{(in)}(Y)G_g(X,Y))\rangle _B
 \ee
   and for baryons
   \be
 G_B(X,Y)=\langle tr(eK^{(f)}\prod^3_{i=1}G_q^{(i)}(X,Y)  eK^{(in)}
 )\rangle _B
 \ee

 The structure of hadron Green's function is clear: it is a product
 of quark and gluon Green's function averaged over the background
 field $B_\mu$. In the next section we shall discuss the way, how to
 make this averaging explicit and express it through the Wilson loop.

 So far we have neglected interaction of propagating quarks and
 gluons with the perturbative field $a_\mu$, i.e. color Coulomb
 exchanges. For quarks those can be easily restored and summed up in
 the exponent [37,38] as we shall show below; for propagating gluons
 perturbative exchanges do not reduce to color Coulomb interaction,
 and should be treated perturbatively term by term [39]. It was
 discussed in [39] that next-to-leading corrections strongly damp the
 main term.

 \section{Relativistic path integrals and relativistic Hamiltonians}

 In this section we shall discuss relativistic path integrals in the
 form of Feynman--Schwinger representation (FSR), first introduced
 for hadron Green's functions in [36] and later developed in
 [37], (for earlier references to FSR see [36-38] and [42]) and
 relativistic Hamiltonians which were obtained from FSR in [4,5].
 These Hamiltonians will be used in the next sections to calculate
 relativistic spectrum of mesons and baryons, and also hybrids and
 glueballs.

 The basic approximation which will be used in this section  is the
 perturbative treatment of spin degrees of freedom, i.e. of fine and
 hyperfine spin interactions. As one will see, for most mesons and
 baryons (except for Nambu--Goldstone mesons $\pi$ and $K$) this is a
 good approximation and the resulting spectra are in good agreement
 with experiment.

 A more general treatment, valid  also for Nambu--Goldstone mesons
 and displaying chiral symmetry breaking, will be discussed in the
 second part of these lectures.

 We start with the meson Green's function (47) and for simplicity
 consider flavour nonsinglet meson, so the last term on the r.h.s. of
 (47) is absent.

 For the quark Green's function in the external field
 $A_\mu=B_\mu+a_\mu$ one can write the FSR [36,37]
\be
G_q(x,y)=(m-\hat D)\int^\infty_0 ds(Dz)_{xy}e^{-K} \Phi_\sigma(x,y)
\ee
where notations used are
\be
K=m^2s+\frac{1}{4}\int^s_0 d\tau(\frac{dz_\mu}{d\tau})^2;~~
D_\mu=\partial_\mu-igA_\mu,
\ee
\be
(Dz)_{xy}=lim_{N\to \infty}\prod^N_{n=1}\frac{d^4\xi(n)}
{(4\pi\varepsilon)^2}\frac{d^4q}{(2\pi)^4} e^{iq(\sum\xi(n)-(x-y))},
N\varepsilon=s
\ee
\be
\Phi_\sigma(x,y)=P_AP_F exp( ig \int^x_y A_\mu dz_\mu)
exp (g\int^s_0
d\tau\sigma_{\mu\nu}F_{\mu\nu}),
\ee
and
$$
\sigma_{\mu\nu}=\frac{1}{4i}(\gamma_\mu\gamma_\nu-\gamma_\nu\gamma_\mu),
~~\xi(n)=z(n)-z(n-1).
$$

One can write the same FSR for the antiquark Green's function; taking
into account that in this case one should use the charge-conjugated
field
\be
A^{(C)}_\mu=C^{-1}A_\mu C=-A^T_\mu,~~F^{(C)}_{\mu\nu}=-F^T_{\mu\nu}
\ee
the ordering of $P_AP_F$ in (55) changes its direction as well as
limits of integration $(x,y)\to (y,x)$. As a result one obtains the
meson Green's function as the path integral over closed Wilson loops
with insertion of magnetic moment terms $\sigma_{\mu\nu} F_{\mu\nu}$
of different signs for quark and antiquark, as it should be,
\be
G_M(x,y)=\langle tr \Gamma^{(f)}(m-\hat D)\int^\infty_0 ds \int^\infty_0
d\bar s e^{-K-\bar K}(Dz)_{xy}(D\bar z)_{xy}\Gamma^{(i)}(\bar
m-\hat{\bar D}) W_F\rangle
\ee
Here the bar sign refers to the antiquark, and
\be
W_F=P_AP_F exp(ig\oint dz_\mu A_\mu) exp(g
\int^s_0
d\tau\sigma^{(1)}_{\mu\nu}F_{\mu\nu})
exp (-g\int^{\bar s}_0\sigma^{(2)}_{\mu\nu}F_{\mu\nu} d\bar\tau)
\ee
In (57) enter integrations over proper times $s, \bar s$ and $\tau,
\bar\tau$, which also play the role of ordering parameter along the
trajectory, $z_\mu=z_\mu(\tau),~~\bar z_\mu=\bar z_\mu(\bar \tau)$.

It is convenient to go over to the actual time $t\equiv z_4$ of the
quark (or antiquark), defining the new quantity $\mu(t)$, which will
play very important role in what follows
\be
2\mu(t) =
\frac{dt}{d\tau},~~ t\equiv z_4(\tau)
\ee
For each quark (or antiquark and gluon) one can rewrite the path
integral (52), (57) as follows (see appendix 1 of the second paper in
[5] for details of derivation)
\be
\int^\infty_0 ds (D^4z)_{xy}... = const \int D\mu(t) (D^3z)_{xy}...
\ee
where $(D^3z)_{xy}$ has the same form as in (54) but with all
4-vectors replaced by 3-vectors, and the path integral $D\mu(t)$ is
supplied with the proper integration measure, which is  derived from
the free motion Lagrangian.

In general $\mu(t)$ can be a strongly oscillating function of $t$ due
to the Zitterbewegung. In what follows we shall use the steepest
descent method for the evaluation of the integral over $D\mu(t)$,
with the extremal $\mu_0(t)$ playing the role of effective or
constituent quark mass. We shall see that in all cases, where spin
terms can be considered as a small perturbation, i.e. for majority of
mesons, $\mu_0$ is positive and rather large even for vanishing quark
current masses $m,\bar m$, and the role of Zitterbewegung is small
(less than 10\% from the comparison to the light-cone Hamiltonian
eigenvalues, see [40,41] for details).

Now the kinetic terms can be rewritten using (59) as
\be
K+\bar K=\int^T_0 dt\{ \frac{m^2}
{2\mu(t)} +\frac{\mu(t)}{2} [(\dot
z_i(t))^2+1]+
\frac{\bar m^2}{2\bar \mu(t)} +\frac{\bar \mu(t)}{2} [(\dot{\bar
z_i}(t))^2+1]\}
\ee
where $T=x_4-y_4$. In the spin-dependent factors the corresponding
changes are
\be
\int^s_0 d\tau \sigma_{\mu\nu} F_{\mu\nu}=
\int^T_0 \frac {dt}{2\mu(t)}\sigma_{\mu\nu} F_{\mu\nu}(z(t)).
\ee
In what follows in this section  we shall systematically do
perturbation expansion of the spin terms, since otherwise (57) cannot
be estimated. Therefore
as the starting approximation we shall use the Green's functions of
mesons made of spinless quarks, which amounts to neglect in (57),(58)
terms $(m-\hat D), (\bar m-\hat {\bar D})$ and
$\sigma_{\mu\nu}F_{\mu\nu}$. As a result one has
\be
G^{(0)}_M(x,y)= const \int D\mu(t) D\bar \mu(t)
 (D^3z)_{xy}
 (D^3\bar z)_{xy}  e^{-K-\bar K}\langle W\rangle .
 \ee

 Our next approximation is the neglect of perturbative exchanges in
 $\langle W\rangle $, which yields for large Wilson loops, $R,T\gg T_g$,
 \be
 \langle W\rangle _B=const~ exp(-\sigma S_{min})
 \ee
 where $S_{min}$ is the minimal area inside the given trajectories of
 quark and antiquark,
 \be
 S_{min} =\int^T_0 dt\int^1_0 d\beta \sqrt{detg},~~
 g_{ab}=\partial_aw_\mu\partial_bw^{\mu},~a,~b=t,~\beta.
 \ee

 The Nambu-Goto form of $S_{min}$ cannot be quantized due to the
 square root and one has to use the auxiliary field approach [42]
 with functions $\nu(\beta,t)$ and $\eta(\beta,t)$ to get rid of the
 square root, as it is usual in string theories. As the result the
 total Euclidean action becomes [5]
 $$
 A=K+\bar K+\sigma S_{min}=
 \int^T_0 dt\int^1_0 d\beta
 \{\frac12(\frac{m^2}{\mu(t)}+\frac{\bar m^2}{\bar
 \mu(t)})+\frac{\mu_+(t)}{2}\dot R^2+
 $$
 \be
 +\frac{\tilde \mu(t)}{2}\dot r^2+
 \frac{\nu}{2}[\dot w^2+(\frac\sigma\nu)^2r^2-2\eta (\dot
 wr)+\eta^2r^2]\}.
 \ee

 Here $\mu_+=\mu+\bar
\mu,~~\tilde\mu=\frac{\mu\bar\mu}{\mu+\bar\mu},~~
 R_i=\frac{\mu z_i+\bar\mu\bar z_i}{\mu+\bar\mu},~~
 r_i=z_i-\bar z_i$. Note, that integrations over $\mu,\nu$ and $\eta$
 effectively amount to the replacement by their extremum values [5].

 Performing Gaussian integrations over $R_{\mu}$ and $\eta$ one
 arrives in the standard way at the Hamiltonian (we take $m=\bar m$
 for simplicity)
$$
H=\frac{p^2_r+m^2}{\mu(\tau)}+
\mu(\tau)+\frac{\hat L^2/r^2}
{\mu+2\int^1_0(\beta-\frac{1}{2})^2\nu(\beta) d\beta}+
$$
\be
+\frac{\sigma^2
r^2}{2}\int^1_0\frac{d\beta}{\nu(\beta)}+
\int^1_0\frac{\nu(\beta)}{2}d\beta,
\label{17}
\ee
where $p^2_r=(\vep\ver)^2/r^2$, and $L$ is the angular momentum,
$\hat L=(\ver\times \vep)$.

The physical meaning of the terms $\mu(t)$ and $\nu(\beta)$ can be
understood when one finds their extremal values. E.g. when $\sigma=0$
and $L=0$, one finds from (67)
\be
H_0=2\sqrt{\vep^2+m^2},~~\mu_0=\sqrt{\vep^2+m^2}
\ee
so that $\mu_0$ is the energy of the quark. Similarly in the limiting
case $L\to \infty$ the extremum over $\nu(\beta)$ yields:
\be
\nu_0(\beta)=\frac{\sigma r}{\sqrt{1-4y^2(\beta-\frac{1}{2})^2}},
~~H^2_0=2\pi\sigma\sqrt{L(L+1)}
\ee
so that $\nu_0$ is the energy density along the string with the
$\beta$ playing the role of the coordinate along the string.

Let us start with the $L=0$ case. Taking extremum in $\nu(\beta)$ one
has
\be
H_1=\frac{p^2_r+m^2}{\mu(t)}
+\mu(t)+\sigma r.
\label{23}
\ee
Here $\mu(t)$ is to be found also from the extremum amd is therefore
an operator in Hamiltonian formalism. Taking extremum in $\mu(t)$ one
obtains
\be
H_{2}=2\sqrt{p_r^2+m^2}+\sigma r.
\label{24}
\ee
The Hamiltonian (71) is what one traditionally exploits in the
relativistic quark model (RQM) [6-8] (apart from color Coulomb and
spin-dependent terms to be discussed below).
The RQM was an essential step in our understanding of hadronic
spectra, see especially [7] for an encyclopedic survey of model
predictions. At the same time the usual input of  RQM contains too
many parameters and the model  was introduced rather ad hoc.  Another
 deficiency of this model at this point is twofold:

i) one usually takes $m$ in (71) to be constituent quark mass of the
order of 100-200 MeV, which is introduced as an input. Instead we
have in (71) the current quark mass renormalized at the reasonable
scale of 1 GeV. Hence it is almost zero for light quarks;

ii) the form (71) is used in RQM for any $L$, and to this end one
writes in (71) $p^2_r \to \vep^2$.

However the Regge slope for both (70) and (71) is $1/8\sigma$ instead
of the string slope $\frac{1}{2\pi\sigma}$,  which occurs for the
total Hamiltonian (69), since the RQM Hamiltonian (71) does not take
into account string rotation.

Still for $L=0$ the form (71) is a good starting approximation, and
it is rewarding that our systematic approach makes here contact with
RQM. Sometimes it is convenient to use instead of (71)a more
tractable form (70) where $\mu(t)= \mu_0$ and $\mu_0$ is to be found
from the extremum of the eigenvalue of Hamiltonian $H_1$:
\be
H_1\psi=M_1\psi,~~M_1=M_1(\mu_0),~~\frac{\partial M_1}{\partial
\mu_0}=0.
\ee
This procedure was suggested in [4], and its accuracy was tested
recently in [43]. In Table 1 listed are eigenvalues of (72) and (71)
$(M^{(1)}_n$ and $M^{(2)}_n$ respectively) for several values of the
radial quantum number $n$.

One can see that the difference between $M^{(1)}_n$ and $M^{(2)}_n$
is around 5\%.

Equation (70), (72) can be reduced to the Airy equation with the
eigenvalues written as (for $m=0$)
\be
M_n^{(1)} =4\mu_0(n),~~ \mu_0(n)=\sqrt{\sigma}(\frac{a(n)}{3})^{3/4}
\ee
where $a(n)$ is the corresponding zero of the Airy function. The
Table 2 gives a set of lowest $a(n)$ and $\mu_0(n)$ for $L=0,1,2$.
The $\mu_0(n)$ for nonzero $L$  values, $L=1,2$ are used later for
spin splittings of obtained levels. The first comprehensive  study of
the equation (70) with fixed $\mu(t)\equiv m$ was done in [44], where
also $a(n)$ are quoted (called there $\zeta (n, L)$).

Finally we turn to the Hamiltonian (67), where the eigenvalues have
been found in [43] using WKB method. The corresponding values are
listed in Table 3 for the  case $m=0$.

An equivalent numerical method was used to find eigenvalues of the
rotating string with quarks at the ends in [45], which should be
compared with the entries of Table 3 and agree with accuracy of about
5\%.
Comparing mass eigenvalues in Table 1-3 one can notice that they
are too heavy to reproduce the experimental values, and one needs to
subtract some 700-800 MeV to get to the region of $\rho$-meson. In
RQM one introduces by hand a large negative constant in the
Hamiltonian, subtracting just this amount.  In our approach here
when spins are treated as a perturbation  we
do not understand how this large negative constant occurs.  Indeed,
there appears a constant $c_0$ due to the finite string width, but
this is too small $c_0\approx 0.1\div 0.2$ GeV to explain the effect.
 It is clear however, that this constant is due to nonperturbative
selfenergies of the quark and antiquark, and should be sensitive to
the quark current mass (i.e.  flavour) and chiral symmetry breaking.
As we shall show  in the second part of these
lectures (to be published separately), the solution of this problem,
which may be called" the Regge intercept problem", lies in the
correct treatment of spin and chiral degrees of freedom, and to
obtain a correct Regge intercept of meson trajectory one should go
beyond the perturbation theory for spin effects.

Generalization of (67) to the case of unequal current masses of quarks is
straightforward [60]
$$
H=\frac{m^2}{2\mu}    +
\frac{\bar m^2}{2\bar \mu} +
\frac{\mu_+}{2} +
\frac{p^2_r}{2\tilde \mu}+
\frac{\hat L^2/r^2}{
2(\mu_1(1-\zeta)^2+\mu_2\zeta^2+\int^1_0
d\beta(\beta-\zeta)^2\nu(\beta))}
 $$
 \be
  +\frac{\sigma^2r^2}{2} \int^1_0
\frac{d\beta}{\nu(\beta)} +\int^1_0 \frac{\nu(\beta)}{2} d\beta,
 \ee
where we have defined
$\zeta=\frac{\mu+\int\beta\nu d\beta}{\mu+\bar \mu+\int \nu d\beta}.$

For $L=0$ one can as before find extremum in $\nu(\beta)$ and obtain
Hamiltonian
\be
H^1=\frac{m^2}{2\mu}    +
\frac{\bar m^2}{2\bar \mu} +
\frac{\mu_+}{2} +
\frac{p^2_r}{2\tilde \mu}+
\sigma r
\ee
 A simplified procedure of solving (75) as in the case of equal mass, (72),
reduces to the solution of Schroedinger--like equation and to the
finding $\mu,\bar \mu$ from the extremum of the total mass [46],
 $$
H^1\Psi= M(\mu,\bar \mu)\Psi,~~ \bar M= M(\mu_0,\bar \mu_), $$ \be
\frac{\partial M}{\partial \mu_0} = \frac{\partial M}{\partial \bar \mu_0} =0
\ee
 here $M(\mu,\bar \mu)$ can be expressed through the same standard numbers
$a(n)$ as given in (73) and Table 2, namely
\be
M(\mu\bar \mu)=
\frac{m^2}{2\mu}    +
\frac{\bar m^2}{2\bar \mu} +
\frac{\mu_+}{2} +
\varepsilon_n(\tilde \mu), ~~
\varepsilon_n(\tilde \mu)= (2\tilde \mu)^{-1/3}\sigma ^{2/3} a(n).
\ee

Extremum conditions for $\mu$ and $\bar \mu$ look like
\be
-\frac{m^2}{2\mu^2} +\frac{1}{2} -\frac{\sigma^{2/3} a(n) \mu^{-4/3}\bar
\mu^{2/3}}{ 2^{1/3}3 (\mu+\bar \mu)^{2/3}}=0
\ee
\be
-\frac{\bar m^2}{2\bar \mu^2} +\frac{1}{2} -\frac{\sigma^{2/3} a(n)
\bar \mu^{-4/3} \mu^{2/3}}{ 2^{1/3}3 (\mu+\bar \mu)^{2/3}}=0 \ee

>From (78), (79) one can see that whenever the current mass is
large, $m\ll \sqrt{\sigma}$, the constituent mass is
close to the current mass, indeed first terms of expansion are \be
\mu^2\cong m^2 [1+\frac{a(n)}{3} (\frac{2\sigma\bar\mu}{m^2(m+\bar
\mu)})^{2/3}] \ee

Till now perturbative gluon exchanges and spin dependent terms in Green's
functions and Hamiltonians have been neglected. Now we are going to restore
their contribution and consider to this end the general form (57).

Leaving out for the moment spin terms $\sigma_{\mu\nu} F_{\mu\nu}$, we have
to do with the Wilson loop containing both perturbative and NP fields,
$W(B+a)$. For the purely perturbative interaction, one can use theorem of
cluster expansion to get
\be
\frac{1}{N_c}\langle tr W(a)\rangle = exp [-\frac{g^2}{2} \int\int dz_\mu dz'\nu
\langle\langle a_\mu(z) a_\nu(z')\rangle\rangle +O(g^4)]
\ee

Introducing perturbative  Green's function (in the Feynman gauge for
simplicity)
\be
\langle\langle a_\mu(z) a_\nu(z')\rangle\rangle =
\frac{\delta_{\mu\nu}}{(2\pi)^2}(z-z')^{-2}
\ee
 and integrating out
$z'_4-z_4$ in $dz_4dz'_4= dz_4 d(z'_4-z_4)$, one obtains the color
Coulomb interaction in the exponent
 \be
  \frac{1}{N_c}\langle tr W(a)\rangle = exp
[ C_2 \alpha_s \int \frac{dz_4}{|\vez-\vez'|}+O_1(\alpha_s)
+O_2(\alpha_s^2)]
\ee
 where $O_1(\alpha_s)$ is the radiative
correction coming from the
spacial  integral $d\vez d \vez'$ in (81), and
$O_2(\alpha_s^2)$ is the corrections coming from quartic and higher
correlators of the form
 $$ \int\int...\int\langle\langle a(1)a(2)a(3)...a(n)\rangle\rangle
dz(1)... dz(n), n\geq 4.  $$

Coming  now  to the original Wilson loop, $W(B+a)$, let us write it in the
form
\be
\frac{1}{N_c}tr\langle  W(B+a)\rangle _{B,a}=
\frac{1}{N_c}tr\langle  [W(B)W(a) W_{int}({B,a)]\rangle _{B,a}}
\ee
where $W_{int}(B,a)$ is the interference term, involving both fields $B_\mu$
and $a_\mu$. Part of this interference was taken into account in the previous
chapter, when the freezing of $\alpha_s$ was considered.
This shows that interference may change behaviour of perturbative terms at
large distances. Another example of interference was considered recently in
[30,31] when treating the small distance behaviour of static
potential.  In both cases it is seen that the main domain of
perturbative fields $a_\mu$  is small distance domain, while that of
NP fields $B_\mu$ is the large distance domain. From the static
$Q\bar Q$ potential one can deduce that both regions join smoothly at
$r_c\approx 0.25 fm$ and perturbative force  is strong for $r<r_c$
(and stronger than both NP and interference), while NP force is
strong for $r>r_c$ and much stronger than both Coulomb force and
interference. So our simplifying assumption is that one can disregard
interference term $W_{int} $ altogether, and take it later as a
correction.

As a result confining term from $W(B)$ and color Coulomb term enter as a sum,
hence it is legitimate to take color Coulomb term as an additional term to be
added to the Hamiltonian (74) or (75). Note that this is true both for heavy
quarks (nonrelativistic situation) and light quarks (relativistic
situation). What is still left over, apart from interference, is radiative
corrections $O(\alpha_s)$ and all higher terms in $O(\alpha_s^2)$. The
reason why the Coulomb term is retained in (83) as compared to those
suppressed, is that Coulomb term is enhanced  as $O(1/v)$ for  slow quarks
and the quadratic term in (81) generates Sudakov--type asymptotics
[38].

As was mentioned above, color Coulomb interaction for the $q\bar q$
system (situaton will be different for gluon systems, see below) is
exponentiated in (83) which implies that Coulomb term can be taken to
all orders, i.e. one should consider the Hamiltonian
\be
H_c=H'-\frac43\frac{\alpha_s(r)}{r}
\ee
where $H'$ is the string Hamiltonian (74) or (75).

One can express the eigenvalue $\varepsilon_n(\tilde \mu)$ of the
$H_c$ through the  eigenvalue of the reduced equation, as was done
before without Coulomb in (73).
\be
\varepsilon_n(\tilde\mu)= (2\tilde\mu)^{-1/3}\sigma^{2/3}
a(\lambda,L,n)
\ee
and the reduced equation looks like [45, 46]
\be
(-\frac{d^2}{d\verho^2}+|\verho|-\frac{\lambda}{|\verho|})\chi(\rho)=
a(\lambda,L,n)\chi(\rho)
\ee
where
\be
\lambda=\frac{4\alpha_s}{3}\frac{(2\tilde
\mu)^{2/3}}{\sigma^{1/3}},~~
r=\frac{\rho}{(2\tilde\mu)^{1/3}},~~
a(\lambda=0, L,n)= a(L,n).
\ee
The eigenvalues $a(\lambda, L, n)$ are given in Table 4 for $L=n=0$.

Finding as before $\mu$ and $\bar \mu$ from the extremum condition on
$M_n$,
\be
M_n=\frac{m^2}{2\mu}+\frac{\mu}{2}+\frac{\bar
m^2}{2\mu^2}+\frac{\bar\mu}{2}+\varepsilon_n(\tilde \mu),
\ee
one obtains equations [46]
\be
1=\frac{m^2}{\mu^2}+\frac{(2\tilde
\mu)^{2/3}\sigma^{2/3}}{3\mu^2}(a(\lambda) +|\frac{\partial
a}{\partial\lambda}|2\lambda)
\ee
\be
1=\frac{\bar m^2}{\bar \mu^2}+\frac{(2\tilde
\mu)^{2/3}\sigma^{2/3}}{3\bar \mu^2}(a(\lambda) +|\frac{\partial
a}{\partial\lambda}|2\lambda)
\ee
one should note, that Coulomb attraction strongly affects the values
of $\mu,\bar \mu$ and wave function at origin.

Connection between eigenfunction $\varphi_n(r)$ of the Hamiltonian
(85) and eigenfunction $\chi_n(\rho)$ of the reduced equation (87) is
\be
\varphi_n(\ver)=(2\tilde \mu\sigma)^{\frac{L}{3}+\frac12}
\chi_n(\rho) Y_{Lm}(\theta,\varphi).
  \ee
  For purely linear
potential, i.e. for $\lambda=0$, the value of the function
$\varphi_n(0)$ does not depend on $n$ [45,47]
\be
|\varphi^{(\lambda=0)}_n(0)|^2=\frac{2\tilde\mu\sigma}{4\pi},~~
\chi^{(\lambda=0)}_n(0)=1.
\ee
Taking into Coulomb potential, i.e.
$\lambda\neq 0$; one gets an enhanced
value of $\chi_n(0)$, which is given
in Table 4. Using these entries and (92)
one can calculate $|\varphi_n(0)|^2$
for any $q\bar q$ system, which will
be used in the next chapter for the
hyperfine interaction and for the
quark decay constants.

For this quantity the asymptotic
freedom is also important and should
be taken into account.

The coefficient
$\rho(AF)=|\frac{\chi^{as}(0)}{\chi(0)}|^2$,
where $\chi^{as}(\rho)$ is the solution of
(87) with $\alpha_s(r)$ in $\lambda$
decreasing according to asymptotic freedom
from the maximal value $\alpha_s^{(0)}$,
with $\Lambda_{QCD}=140 MeV$ is given in
 Table 5.

 Now we turn to the spin terms, considering  them as perturbation
(a complete treatment of spin terms within new formalism will be
given in the last section) In the derivation of spin--dependent
interaction valid also for massless quarks the advantage of the
present method will be evident, since the usual treatment exploits
expansion in inverse powers of quark masses, appropriate for heavy
quarkonia only.  In our case the problem reduces to the calculation
of averages in (57), (58), with the use of (62). This can be written
as [48,49]
 \be \langle W_F\rangle =P_F exp (\int^T_0
\frac{dt}{2\mu(t)}\sigma^{(1)}_{\mu\nu} \frac{\delta}{i\delta
s_{\mu\nu} (z(t))}) exp (-\int^T_0\frac{d\bar t}{z\bar \mu(\bar
t)}\sigma^{(2)}_{\mu\nu} \frac{\delta}{i\delta s_{\mu\nu} (\bar
z(\bar t))}) \langle W\rangle  \ee where for $\langle W\rangle $ one can use the cluster
expansion (28).

In the Gaussian approximation one has $(P_F\to P_\sigma)$
\be
\langle W_F\rangle =P_\sigma exp(K_1+K_2+K_{11}+K_{22}+K_{12})\langle W\rangle
\ee
and e.g.
   \be
K_1=ig^2\int^T_0\frac{dt}{2\mu(t)}\sigma^{(1)}_{\mu\nu}\int
ds^{(w)}_{\lambda\rho}\langle F_{\mu\nu} (z(t))F_{\lambda\rho}(w)\rangle ,
\ee
   \be
K_{12}=-g^2
\int^T_0\frac{dt}{2\mu(t)}
\int^T_0\frac{d\bar t}{2\bar \mu(\bar t)}
\sigma^{(1)}_{\mu\nu}
\sigma^{(2)}_{\lambda \rho}
\langle F_{\mu\nu} (z(t))F_{\lambda\rho}(\bar z(\bar t))\rangle ,
\ee
while other terms are obtained by replacing
$\sigma^{(1)}_{\mu\nu}
\frac{dt}{2\mu(t)} $
by
$\sigma^{(2)}_{\mu\nu}
\frac{d\bar t}{2\bar \mu(\bar t)} $ or vice versa. In what follows we
do not need the selfenergy terms $K_{11}, K_{22}\sim
(\sigma^{(i)}_{\mu\nu})^2$.

Now  one use representation (33) to express field correlators
through $D(u), D_1(u)$. To define the minimal surface for integration
in $\langle W\rangle , \langle W_F\rangle $ one connects trajectories of quark $z(t)$ and
antiquark $\bar z(\bar t)$ by a smooth line $w(t,\beta)$ at $t=\bar
t, 0\leq \beta \leq 1$. One can approximate it by  a straight line
for simplicity: $w(t,\beta)=z(t)\beta+\bar z(t) (1-\beta)$.

Introducing angular momenta in Minkowski space
\be
L_i^{(1)}= (\ver\times \vep_1)_i= ie_{ikm}r_k(t)\mu\dot z_m(t),
\ee
\be
L_i^{(2)}= (\ver\times \vep_2)_i= ie_{ikm}r_k(t)\bar \mu\dot{\bar
z}_m(t), \ee
 one has
  \be
  ds_{ik}=(dw_i\dot w_k-dw_k\dot w_i)=
  \frac{1}{i}d\beta dt e_{ikm} (\frac{\beta L_m^{(1)}}{\mu} +
  \frac{(-\beta)L_m^{(2)}}{\bar \mu})
  \ee
  We refer the reader for details to [48-50].

  The resulting spin terms can be written in the Eichten--Feinberg
  notations [51], namely
  $$
  V_{SD}(r)=(
  \frac{\sigma_i^{(1)}L_i^{(1)}}{4\mu^2}-
  \frac{\sigma_i^{(2)}L_i^{(2)}}{4\bar
  \mu^2})
  (\frac{1}{r}\frac{d\varepsilon}{dr}+
  \frac{2}{r}\frac{dV_1}{dr})+
  $$
  \be
  \frac{\sigma_i^{(1)}L_i^{(1)}
  -\sigma_i^{(2)}L_i^{(2)} }
  {2\mu\bar \mu}
  \frac{1}{r}\frac{dV_2}{dr}+
  \frac{\sigma_i^{(1)}\sigma_i^{(2)}}{12\mu \bar
  \mu} V_4(r)+
  \ee
  $$
  \frac{1}{12\mu\bar \mu}(3
  \sigma_i^{(1)}n_i \sigma_k^{(2)}n_k
  -\sigma_i^{(1)}\sigma_i^{(2)})V_3(r)
  $$
  and $V_i(r), i=0,1,2,3,4$ are expressed through $D,D_1$ [48,50] as
  follows.
  \be
\frac{1}{r}\frac{dV_1}{dr}=-\int^{\infty}_{-\infty}d\nu\int^r_0
\frac{d\lambda}{r}
(1-\frac{\lambda}{r})D(\lambda,\nu)
\label{05}
\ee
\be
\frac{1}{r}\frac{dV_2}{dr}=\int^{\infty}_{-\infty}d\nu\int^r_0
\frac{\lambda d\lambda}{r^2}
[D(\lambda,\nu)+D_1(\lambda,\nu)+\lambda^2\frac{\partial
D_1}{\partial\lambda^2}]
\label{06}
\ee
\be
V_3=-\int^{\infty}_{-\infty} d\nu r^2\frac{\partial
D_1(r,\nu)}{\partial r^2}
\label{003}
\ee
\be
V_4=\int^{\infty}_{-\infty}d\nu
(3D(r,\nu)+3D_1(r,\nu)+2r^2\frac{\partial D_1}{\partial r^2})
\label{08}
\ee
\be
\frac{1}{r}\frac{d\varepsilon
(r)}{dr}=\int^{\infty}_{-\infty}d\nu\int^r_0 \frac{
d\lambda}{r}
[D(\lambda,\nu)+D_1(\lambda,\nu)+(\lambda^2+\nu^2)\frac{\partial
D_1}{\partial\nu^2}]
\label{09}
\ee
 Equations (93-97) give the NP spin-dependent interaction through
 functions $D, D_1$; the latter have been measured on the lattice [22
 -
 24] to be of the form
 \be D(x) = D(0) exp (-\frac{|x|}{T_g}),~~
D_1(x) = D_1(0) exp (-\frac{|x|}{T_g}) \ee
with
$\frac{D_1(0)}{D(0)}=\frac13(\frac17)$, and $T_g=0.2(0.33)fm$ for
quenched [22,23] (dynamical [24]) quarks; while $D(0)=0.073 GeV^4$
for quenched case [22].

Insertion of (107) into (102-106) yields an estimate of all 5
potentials $\varepsilon (r), V_i(r),i=1,2,...4$. The first,
$\varepsilon (r)$ is just scalar potential $V(r)$ appearing in (37),
and present in  (75) in the form $\sigma r; V_3(r)$ and $V_4(r)$ are
numerically small as compared to the perturbative contributions to be
discussed below. The largest contribution comes from the spin--orbit
term; $V_{so}$; since $V'_1(r)|_{r\to\infty} = -\varepsilon'
(r)|_{r\to \infty} =-\sigma$, one obtains for the equal mass case the
asymptotics $ (L_i^{(1)}=-L_i^{(2)}= L_i)$ \be
V_{so}(r)=-\frac{\sigma \veS\veL}{2\mu r}f(r),~~ f(r\to\infty)=1.
\ee
For small $r$ the behaviour of $f(r)$ is changed, $f(0)=0$, and
$V_{so}(r\to o)=const$. For more details and discussion of the
behaviour of $v_i(r)$ and comparison with lattice data see [50].

Analysis of spectra of heavy quarkonia made in [32] strongly prefer
$f(0)=1$, in contradiction with (101-103). The way out may be
seen in the interference terms at small $r$, studied in [30,31], which
show that linear dependence of $\varepsilon(r)$ in $r$ may persist at
very small $r$ in agreement with a  recent dedicated lattice
analysis [14].

We turn now to the perturbative contribution to spin--dependent
interaction and to this end one can use again the general expressions
(101-106) where instead of $D,D_1$ one should insert the
corresponding perturbative expressions. To the order $O(g^2)$ one has
\be
D(x)=0,~~ D_1(x)=\frac{16 \alpha_s}{3\pi x^4}
\ee
Insertion of (109) into (101-106) yields
\be
V_{1p}=0;~\frac1R
V'_{2p}(r)=\frac{4\alpha_s}{3R^3},~
V_{3p}=\frac{4\alpha_s}{R^3},~
\varepsilon_{p}(R)=-\frac{4\alpha_s}{3R},~
V_{4p}(R)=\frac{32\pi\alpha_s}{3}\delta^{(3}(\veR).
\ee
The final expression for the mass of the meson with spin interaction
taken into account perturbatively is obtained from (75) and (85),
(101) and has the form $$ M_n=\bar M_n(\mu_0,\bar \mu_0)+\frac{32
\ves_1\ves_2\pi\alpha_s}{9\mu_0\bar\mu_0}\varphi^2_n(0)+
$$
$$
(\frac{\ves_1}{2\mu^2_0}+
\frac{\ves_2}{2\bar \mu^2_0})\veL\langle \frac{-\sigma f(r)}{r}\rangle +
(\frac{\veS}{\mu_0\bar \mu_0}+
\frac{\ves_1}{2\bar \mu^2_0}+
\frac{\ves_2}{2\bar \mu^2_0})\veL\frac43\langle \frac{\alpha_s}{r^3}\rangle +
$$
\be
\frac{\langle 3\ves_1\ven\ves_2\ven-\ves_1\ves_2\rangle }{\mu_0\bar\mu_0}\frac43
\langle \frac{\alpha_s}{r^3}\rangle
\ee
Where $\veS=\ves_1+\ves_2$,
and
\be
\bar M_n(\mu_0,\bar
\mu_0)=\frac{m^2}{2\mu_0}+\frac{\mu_0}{2}+\frac{\bar
m^2}{2\bar\mu_0}+\frac{\bar\mu_0}{2}+\varepsilon(\tilde\mu_0)+\Delta
M-C_0\equiv M_n^{(0)}+\Delta M-C_0
\ee

Here $\Delta M$ takes into account the so-called string correction,
i.e. the difference of eigenvalues of the Hamiltonian (67) and (70)
due to the string rotation. According to [5] this can be written as
\be
\Delta M=-\frac{16\sigma^2L(L+1)}{3(M_n^{(0)})^3}.
\ee
Let us note the main properties of the mass formula (111). First of
all it has the minimal possible number of input parameters; those are
{\bf current} quark masses (defined at the appropriate scale of $1
GeV$), string tension, $\alpha_s(r)$ or equivalently $\Lambda_{QCD}$
and the overall subtraction constant $C_0$.

We {\bf  do not introduce constituent quark masses}
and do not expand in inverse quark masses to get spin interaction, as
it is common in relativistic quark models. In addition we take into
account string rotation, which is not done in those models.

Moreover, our expression of the total mass $M_n$ is the result of
derivation from the first principles of QCD with all steps and
approximations clearly visible, so that one can check and improve if
need be. This is in contrast to the ad hoc model building principle.

Now we come to the confronting of our approach and first of all mass
formula (111) with experiment and lattice data.

\section{Spectrum of light mesons, hybrids and glueballs}

\subsection{}

The spectrum of mesons and glueballs (and of any two constituents
connected by the string with  spins and Coulomb force neglected) is
given by the Hamiltonian (74).

The denominator in (74) is actually the moment of inertia of the
system, comprising both quark (the $\mu$ term) and string (the $\nu$
term) contributions.

The eigenvalues of (74) have been  found quasiclassically in [44] and
given in Table 3  for the case of zero--mass quarks. For the case of
two--gluon glueballs the corresponding eigenvalues are obtained by
multiplication with the square root of Casimir factors ratio; since
for the adjoint string $\sigma_{adj}=
\sigma_{fund}\frac{C_2(adj)}{C_2(fund)}=\frac94 \sigma_{fund}
$ In Fig. 1 shown are trajectories of eigenvalues of (74) as
functions of $L$ and $n$. The Regge slope is close (within 3\%) to
the  string slope  $(2\pi\sigma)^{-1}$. Note also the linear
trajectories in $n$ with approximately 2 times smaller slope.

The resulting spectrum for light--light mesons
was found first in [34] and  is close in basic
features to that of Isgur  and Godfrey [7] for $L=0$, but differs
for $L>0$, since we take into account string rotation, and differs
somewhat in spin splittings, since we have used for
spin--dependent (SD) terms expressions derived from QCD (102-106),
(110). In Fig.1 also shown
is the experimental $\rho$ type Regge trajectory, which has almost
the same slope, but shifted downwards. For this reason we introduce
at this point (as in most models) the negative constant $C_0$, as a
surrogate of the nonperturbative selfenergy term for each quark,
while we do not introduce such term for the valence gluon. The
reason for this procedure and an estimate of this selfenergy will
be given in the second part of lectures (published separately),
 where effects of chiral symmetry breaking are taken into
account.

An example of the calculated spectrum of light--light mesons is
given in Table 7 for $I=1$ case and compared to experiment. It is
seen that the overall agreement is reasonable and comparable to
that of Isgur--Godfrey model [7]. Note, however, that in [7] there
are 14 fitting parameters, while we have in our approach only 3:
$\alpha_s$ , $\sigma$ and $C_0$. The most important fact is that
spin--dependent potentials are exactly derived with masses
$\mu_0,\bar \mu_0$ without any expansion in inverse mass powers
assumed in all potential quark models; moreover $\mu_0,\bar \mu_0$
are computed and expressed through $\sqrt{\sigma}$.

Several words should be said about  pion. The present approach
as a whole
is based on the assumption that spin
corrections can be computed as a perturbation. This also implies
smallness of chiral symmetry breaking effects. Now for the pion this
is not valid, hyperfine correction is too large, and what is even
worse, if one tries to treat the spin-spin term
$-\frac{8\pi\alpha_s}{3\mu^2_0}\delta^{(3)}(\bar r)$
to higher orders, then the minimization of the mass $M_n$ (111) in
$\mu_0$ is no more possible, since at $\mu_0\to 0$ the spin-spin term
diverges. This happens only in the pionic channel and signals the
instability of the vacuum and the onset of chiral symmetry breaking.
Therefore pion should not  be considered in the framework of the
approach described.

For radially excited pion states the spin-spin correction is smaller
and the present approach may be more reasonable. Some of these states
are presented in Table  7.

\subsection{Hybrids}

Hybrids and glueballs are the most interesting examples where all
advantages of our approach can be seen  clearly. First  of all
this is the definite construction of the hybrid ( and glueball)
Green's function (49) based on the background perturbation theory.
 Note the difference of our definition of hybrids from that of the
 flux-tube model; at the same time the structure of in and out
 hybrid states $\Psi_{in,f}$ is similar to that of the lattice QCD.
 Moreover, since for hybrids we do not introduce any new parameters
 (and effective mass of the gluon $\mu_g$ will be calculated through
 string tension), one can predict hybrid and glueball masses
 unambiguously, in this way  seriously checking the whole formalism.
The first studies of hybrid spectra in this method have been done in
[52-54]. For later development using variational methods see [55].

 Let us turn to the construction of the hamiltonian for the hybrid
 system. We start with the hybrid Green's function and treating spin
 terms as perturbation, disregard in the first approximation the
 gluon spin term $2ig F_{\mu\nu}$ in (9). Using FSR we get similarly
 to (57) the form
 $$
 G_{hyb}(x,y)=\langle tr
 \Gamma^{(f)}(m-\hat D)
 \int^\infty_0 ds
 \int^\infty_0 d\bar s
 \int^\infty_0 ds_g
 e^{-K-\bar K-K_g}
 (Dz)_{xy}
 (D\bar z)_{xy}
 (Du)_{xy}
 $$
 \be
 \times
 \Gamma^{(i)}(\bar m-\hat{\bar D}) W_{hyb}\rangle
 \label{h.1}
 \ee
 where
 \be
 W_{hyb}=(\Phi(x,y))_{\alpha\beta} t^a_{\beta\gamma} \hat \Phi_{ab}
 (x,y) t^b_{\delta\alpha} \Phi_{\gamma\delta} (y,x),
 \label{h.2}
 \ee
 and $\Phi(x,y), (\Phi (y,x))$ are transporters (27) belonging to the
 quark (antiquark) Green's function, while $\hat \Phi(x,y)$ is that
 of the gluon in the adjoint representation. At this point one can
 use the large $N_c$ approximation, which yields accuracy around 10\%
 in the most problems of QCD.

 Then the gluon line can be replaced by the 'tHooft rule with a
 double $q\bar q$ line,
 \be
 t^a_{\beta\gamma}\hat \Phi_{ab} t^b_{\delta\alpha} \to
 \Phi_{\beta\alpha} (x,y) \Phi_{\gamma\delta } (y,x)
 \label{h.3}
 \ee
 and the vacuum average of $W_{hyb}$ reduces to the product
 \be
 \langle W_{hyb}\rangle =\langle W(C_1)\rangle \langle W(C_2)\rangle
 \label{h.4}
 \ee
 where $C_1, C_2$ are adjacent contours made of quark (antiquark)
 trajectory and gluon trajectory.

 The subsequent transformations with (\ref{h.1}) are the same as in
 the $q\bar q$ case leading from (57) to (67).

 We shall use only the simplest form of the  Hamiltonian for zero
 angular momenta, equivalent to (70)
 and (112). In the case of the hybrid state
 it is
 \be
 H_{hyb}=
 \frac{m^2_1}{2\mu_1}+
 \frac{m^2_2}{2\mu_2} +
 \frac{\mu_1+\mu_2+\mu_3}{2}+h_H-C_0(q\bar q g)
 \label{h.5}
 \ee
 with
 \be
 h_H=-\frac{1}{2m}(\frac{\partial^2}{\partial \vexi^2}+
 \frac{\partial^2}{\partial\veta^2})+
\sigma|\vez^{(1)}-\vez^{(3)}|+
\sigma|\vez^{(2)}-\vez^{(3)}|;
\label{h.6}
\ee
where we have defined c.o.m. coordinate and total effective mass
respectively as
\be
R_i=\sum^3_{k=1}\frac{\mu_kz_i^{(k)}}{\mu},~~
\mu = \sum^3_{k=1}\mu_k
\label{h.7}
\ee
while Jacobi coordinates $\xi_i, \eta_i$, are expressed through
those of $q,\bar q$:  $z_i^{(1)}, z_i^{(2)}$ and of gluon $z^{(3)}_i$
as
$$
z_i^{(3)}= R_i-\sqrt{\frac{m(\mu_1+\mu_2)}{\mu\mu_3}}\xi_i
$$
$$
z_i^{(1)}= R_i
+\sqrt{\frac{m\mu_3}{\mu(\mu_1+\mu_2)}}\xi_i
+\sqrt{\frac{m\mu_2}{\mu_1(\mu_1+\mu_2)}}\eta_i
$$
\be
z_i^{(2)}= R_i
+\sqrt{\frac{m\mu_3}{\mu(\mu_1+\mu_2)}}\xi_i
-\sqrt{\frac{m\mu_1}{\mu_2(\mu_1+\mu_2)}}\eta_i
\label{h.8}
\ee

 Here $m$ is an arbitrary mass parameter.
At this point one should comment on gluon exchanges between the
valence  gluon in the hybrid and quark (or antiquark). As we shall
see in the next section, devoted to glueballs, a valence gluon does
not have a color Coulomb interaction with another gluon or quark: the
OGE diagrams existing in the lowest order do {\bf not} sum up in the
ladder--type series of the Coulomb interaction, since exchanged gluon
and valence gluon are identical and moreover four--gluon vertex is
also important. This is in contrast to the $q\bar q$ or $qqq$
systems, where the Coulomb series occurs naturally as a first term in
the cluster expansion series, see (81). The perturbative series for
the gluon--gluon case is summed up in the Lipatov approach [56] and
the result for the gluon ladder is strongly damped by the  one--loop
corrections. Therefore we disregard here the corresponding Coulomb
terms for $qg$ and $\bar q g$ systems.

The Hamiltonian (118) can be used in problems of two types. First, one
can consider, as it is done on the lattice, [57], fixed (very heavy)
quark and antiquark at a distance $R$ from each other, and gluon in
some angular momentum and spin state.

In this case one can define as in [57] the potentials $V_{r}(r)$ for
the static $Q\bar Q$ pair at distance $r$ with gluon in a state,
which can be denoted as in the diatomic molecule, i.e. with
projection of the total gluon momentum $\veJ$ on the axis $\ver$,
\be \Lambda= \veJ\ver/r,~~\Lambda =0,1,2,... (\Sigma,\Pi,\Delta,
\Phi) \label{h.9}
 \ee
  and combined operator of inversion w.r.t.
midpoint of $Q\bar Q$ and charge conjugation $\eta_{cp}=1(-1)$
denoted as $g(u)$. $\Sigma$ can be also even (odd) w.r.t reflection
in the plane of $Q\bar Q$, denoted as $+(-)$.

The Hamiltonian for the gluon in such two--center problem looks like
\be
H_{Q\bar Q g}-H_{Q\bar Q}=
\frac{\mu_3}{2}+
\frac{\vep^2}{2\mu_3}+
\sigma|\vez^{(3)}-\vez_Q|+\sigma|
\vez^{(3)}-\vez_{\bar Q}|+V_{LS}+\frac32\frac{\alpha_s}{r}
\ee
where the spin-orbit term is similar to that in (101), while the
Coulomb term is a result of subtraction from the coulor-octet
repulsion $\frac{\alpha_s}{6r}$ in the $Q\bar Qg$ system of the
attractive color--singlet term $-\frac{4\alpha_s}{3r}$.

The Hamiltonian (123) can be studied numerically.

A rough estimate of the eigenvalues of (123) is obtained when one
expands string potentials in (123) assuming that gluon slightly
vibrates around the middlepoint of $Q\bar Q$, expanding the square
roots.

Then the spectrum is
\be
\Delta V_{Q\bar Q g} (r) =
(\frac{4\sigma}{r})^{1/3} \left(n+\frac32\right)^{2/3}
+\frac32\frac{\alpha_s}{r},~~ n=0,1,2...  \label{h.11}
 \ee
 A comparison of (124) with the lattice calculation [57]
  shows a good  qualitative agreement.

We turn now to the real physical objects, $q\bar q g$ systems with
light or heavy quarks. In this case one should take quark kinetic
energies into account and consider $q\bar q g$ as a 3--body system
with the Hamiltonian given in (118). The most efficient way to treat
this problem is the hyperspherical expansion [58]. One defines the
hyperradius $\rho$ as
\be
\rho^2=\eta^2+\xi^2,
\label{h.12}
\ee
and expands the wave function in a series
\be
\Psi(\vexi, \veta)=\sum_{K\nu n}\chi_{KN}(\rho) \Omega_{K\nu}
\label{h.13}
\ee
where $\Omega_{K\nu}$ is hyperspherical angular function [58], and
$\nu$ denotes all quantum  numbers in addition to the  global
momentum $K=0,1,2,...$

Keeping only one harmonics with the given $K$, one obtains radial
equation (neglecting for the moment Coulomb and spin--dependent
terms)
\be
[-\frac{d^2}{2md\rho^2}+W(\rho)] \chi_{Kn}(\rho)=
\varepsilon_{Kn}\chi_{Kn}(\rho)
\label{h.14}
\ee
where
\be
W(\rho)=\frac{\alpha(\alpha+1)}{2m\rho^2}+
\sigma\rho\frac{32}{15\pi}  (\alpha_{13}+\alpha_{23})
\label{4.15}
\ee
and
\be
\alpha=K+\frac{3}{2},
~~\alpha_{i3}=\sqrt{\frac{m(\mu_i+\mu_3)}{\mu_i\mu_3}},~~
i=1,2.
\label{h.16}
\ee
For an estimate with  some 5\% accuracy, one can find
$\varepsilon_{kn}$ by simply minimizing $W(\rho)$ in $\rho$.

Assuming now equal quark masses, $m_1=m_2$ one could obtain
$\mu_1=\mu_2$, and taking the total hybrid mass as
\be
M_{Kn}=\frac{m^2}{\mu_1}+\mu_1+\frac{\mu_3}{2}+\varepsilon_{Kn}
-C_0(q\bar q g)
\label{h.17}
\ee
and finding the minimal value $\varepsilon_{Kn}$ from $W(\rho_0)$,
one has
\be
M_{Kn}
=\frac{m^2_1}{\mu_1}+
\mu_1+\frac{\mu_3}{2}+\frac32\frac{\sigma^{2/3}}{\mu_{13}^{1/3}} \bar
C
-C_0(q\bar q g)
\label{h.18}
\ee
where
$\bar
C=(\frac{64}{15\pi})^{2/3}[\alpha(\alpha+1)]^{1/3}$. It is
reasonable to take $C_0(q\bar qg)$ the same as in the $q\bar q$ case
since the constant term is due to quark selfenergy terms only.

In the limit of heavy mass $m_1\gg \sqrt{\sigma}$, one
get from minimization $\mu_1\approx  m_1$, and
minimizing in $\mu_3$ one obtains
\be
M_{Kn}=2m_1+2\sqrt{\sigma}\bar C^{3/4}-
C_0(q\bar q g)
\label{h.19}
\ee

In particular, subtracting the heavy mass $2m_1-C_0$
one gets
\be
\Delta M= 2\sqrt{\sigma} \bar C^{3/4} =
\begin{array}{l}
2.72\sqrt{\sigma}=1.375 GeV, ~~K=0\\
3.44\sqrt{\sigma}=1.7 GeV, ~~K=1
 \end{array}
 \label{h.20}
 \ee
 For light quarks one takes the same $C_0$ for a light hybrid as for
  the corresponding light meson.
   For the lowest exotic hybrid
 $1^{-+}$ one should take $K=1$ ($L=1$ for the gluon) and agreement
 with recent lattice calculation of the $b\bar b g$ mass excitation is
 very good, see Table 8.

\subsection{Glueballs}

This section is based  on papers [59,60].
  The $L=0$ Hamiltonian for glueballs (neglecting spin and
  perturbative interaction) is obtained from that of $q\bar q$ system
  by replacing $\sigma_{fund}$ by $\sigma_{adj}$.

\be
H'_0= \frac{\vep^2}{\mu_0} +\mu_0+\sigma _{adj}r
\ee
The value of $\sigma_{adj}$ in (134) can be found from the string
tension of $q\bar q$ system, since the Casimir scaling found on the
lattice [33] predicts that
\be
\sigma_{adj}= \frac{C_2(adj)}{C_2(fund)} \sigma_{fund} =\frac94
\sigma_{fund }
\ee
For light quarks the value of $\sigma_{fund}$ is found from the slope
of meson Regge trajectories and is equal to
\be \sigma_{fund} =\frac{1}{2\pi\alpha'}\approx 0.18 GeV^2
\ee
>From that we find
\be
\sigma_{adj}\approx 0.40 GeV^2
\ee

In what follows the parameter $\mu$ and its optimal value $\mu_0$,
which enters in (134) play very important role. The way they enter
spin corrections in (101) and magnetic moments shows that $\mu_0$
plays the role of effective (constituent) gluon mass (or constituent
quark mass in the equation for the $q\bar q$ system).

In contrast to the potential models, where the constituent mass of gluons and
quarks is introduced as the fixed input parameter in addition to the
parameters of the potential, in our approach $\mu_0$ is calculated from the
extremum of the eigenvalue of equation (134), which yields
$$
\mu_0(n)=\sqrt{\sigma}(\frac{a(n)}{3})^{3/4},
M_0(n)=4\mu_0(n)
$$
where $\sigma=\sigma_{adj}$ for gluons and $\sigma=\sigma_f$ for masses
quarks, and a(n) is the eigenvalue of the reduced equation
$\psi^{''}+(a(n)-\rho-L(L+1)/\rho^2)\psi=0$
The first several values of a(n) and $\mu_0(n)$ are given in the Table
2, and
will be used below.

Note that our lowest "constituent gluon mass" $\mu_0(n=L=0)=0.528
GeV$ (for $\sigma_f=0.18 GeV)$ is rather close to the values
introduced in the potential models, the drastic difference is that
$\mu_0$ depends on $n,L$ and grows for higher states.

>From Table 1,2,3 and Fig. 1 one can see that mass spectra of the
Hamiltonian (67) are described with a good accuracy by a very simple
formula (a similar conclusion follows from calculations in [45])
\be
\frac{M^2}{2\pi\sigma}=L+2n_r+c_1
\ee
where $L$ is the orbital momentum, $n_r$ --radial quantum number
and $c_1$ is a constant $\approx 1.5$. It describes an infinite set
of linear Regge-trajectories shifted by $2n_r$ from the leading one
($n_r=0)$.
The only difference between light quarks and gluons is the value of
$\sigma$, which determines the mass scale.

Thus the lowest glueball state with $L=0, n_r=0$ according to eqs.
(6), (5) has $M^2=4.04 GeV^2$.

It corresponds to a degenerate $0^{++}$ and $2^{++}$ state.
\be
M=2.01 GeV
\ee

In order to compare our results with the corresponding lattice calculations
[61-63] it is convenient to consider the quantity $\bar
M/\sqrt{\sigma_f}$, which is not sensitive to the choice of string
tension $\sigma$\footnote[2]{Note that the value $\sigma_f\simeq
0.21 GeV^2$ used in lattice calculations differs by about 20\% from
the "experimental" value $\sigma_f=0.18 GeV^2$.}. The spectrum of
glueball states obtained in lattice calculations is given in the
Table 9, where masses of glueballs for
values of $\sigma_f$ used in these calculations are given.

>From these data we have for $L=0, n_r=0$ states the spin averaged mass
\be
\frac{\bar
M}{\sqrt{\sigma_f}}=\frac{M(0^{++})+2M(2^{+})}{3}\frac{1}{\sqrt{\sigma_f}}
\ee
the value $4.61\pm 0.1$, which should be compared to our prediction
$\bar M^{theor}(L=0, n_r=0)/\sqrt{\sigma_f}=4.60.$

For radially excited states our theory predicts
\be
\frac{\bar M^{theor}}{\sqrt{\sigma_f}}(L=0,n=1)=7.0
\ee

Lattice data [61] give for thise quantity
\be
\frac{\bar M^{lat}}{\sqrt{\sigma_f}}(L=0,n_r=1)=6.8\pm 0.35
\ee

For $L=1, S=1$ states one can define spin-averaged mass in a similar
way
\be
\frac{\bar M}{\sqrt{\sigma_f}}=\frac{M(0^{-+})+2M(2^{-+})}{3}\frac{1}{\sqrt{\sigma_f}}
\ee
 lattice data [61-63] yield
\be
\frac{\bar M^{lat}}{\sqrt{\sigma_f}}(L=1,n=0)=6.36\pm 0.5;
\frac{\bar M^{lat}}{\sqrt{\sigma_f}}(L=1,n=1)=8.1\pm 0.5
\ee
which is in a reasonable agreement with our prediction
\be
\frac{\bar M^{theor}}{\sqrt{\sigma_f}}(L=1,n_r=0)=5.95;
\frac{\bar M^{theor}}{\sqrt{\sigma_f}}(L=1,n_r=1)=8.0
\ee

For $ L=2,n_r=0$ the spin averaged state has
\be
\frac{\bar M^{theor}(L=2,n_r=0)}{\sqrt{\sigma_f}}=7.0
\ee

Lattice data [61] exist only for $3^{++}$ state which yields
$\frac{M(3^{++})}{\sqrt{\sigma_f}}=7.7$.  The overall comparison of
spin-averaged masses computed by us and on the lattice shows  a
striking agreement.

Thus we come to the conclusion that the spin--averaged masses
obtained from purely confining force with relativistic kinematics for
valence gluons are in a good correspondence with lattice data, which
implies that the dominant part of glueball dynamics is due to
the QCD strings.

   \subsection{Spin splittings of glueball masses}

   Here we shall treat spin effects in a perturbative  way, in the
   same manner, as it is done with spin effects in heavy
   and  light quarkonia; a
   glance at the lattice results given in Table 9 tells that spin
   splittings in glueball states  apart from $2^{++}-0^{++}$ amount to less
than 10-15\% of the total mass, and hence perturbative treatment is justified
   to this accuracy level.

   The two--gluon mass operator can be written as
   \be
   M=M_0(n,L) +\veS\veL M_{SL}+\veS^{(1)}\veS^{(2)} M_{SS}+M_T,
   \label{7.1}
   \ee
   where $M_0$ is the eigenvalue of the Hamiltonian $H'\equiv
   H_0+\Delta H_{pert}$, and $H_0$ is given in (134),
   while $\Delta H_{pert}$ is due to
   perturbative gluon exchanges and as discussed in [60], gives a
   small correction, which will be omitted here.

   To obtain three other terms in (147) one should consider averaging
   of the operators $\hat F$ in (9)
   which enter in  exponent for the Green's function  and take into
   account that
   \be
   -2i\hat F_{\mu\nu}=2(\veS^{(1)} \veB^{(1)}+ \tilde{\veS}^{(1)}
   \veE^{(1)})_{\mu\nu}
   \label{7.2}
   \ee
   and similarly for the term in the integral $\int\hat F d\tau'$,
   with the replacement of indices $1\to 2$.
Here gluon spin operators are introduced, e.g.
\be
(S^{(1)}_m)_{ik}=-ie_{mik},~~i,k =1,2,3, (\tilde S^{(1)}_m)_{i4}
=-i\delta_{im}
\label{7.3}
\ee
Background gauge condition (46) allows to exclude $\mu=\nu=4$, and
hence $\tilde S^{(1)}$ in  (148). Since the structure of the term
$\hat F$  is the same as in case of heavy quark
with the replacement of the heavy quark mass  by the effective gluon
parameter $\mu_0$(see (134)), one can use the spin analysis of
heavy quarkonia done in [46,50], to represent the spin--dependent
part of the Hamiltonian in the form similar to that of Eichten and
Feinberg [51]
 $$ \Delta H_s= \frac{\veS\veL}{2\mu^2_0} (\frac2r
\frac{dV_1}{dr} +\frac2r\frac{dV_2}{dr})+
   \frac{\veS^{(1)}\veS^{(2)}}{3\mu^2_0} V_4 (r)+
   $$
   \be +
   \frac{1}{3\mu_0^2} (3(\veS^{(1)}\ven)(\veS^{(2)}\ven)
   -\veS^{(1)}\veS^{(2)})V_3(r)+\Delta V
   \label{7.4}
   \ee
   where $\veS=\veS^{(1)}+\veS^{(2)},\Delta V$ contains higher
   cumulant contributions (which can be estimated to be   of the
   order of 10\% of the main term in (150)).

   The functions $V_i(r)$ are the same as   for heavy quarkonia
   [48,50] except that
   Casimir operator is that of adjoint charges,
    the corresponding expressions of
   $V_i(r)$ in terms of correlators $D(x), D_1(x)$, are given in
   (46,50). Both $D$ and $D_1$ are measured on the lattice [22-25]
   and $D_1$ is found to be much smaller  than $D$. Therefore one can
   neglect the nonperturbative part of $V_3(r)$, while that of $V_4$
   turns out to be also  small numerically, $M_{SS}(nonpert.) < 30
   MeV$, and we shall also neglect it.

   The only sizable spin--dependent nonperturbative contribution
   comes from the term $\frac{dV_1}{dr}$ (Thomas precession) and can
   be written asymptotically at large $r$ as
   \be
   \Delta H(Thomas) =
   -\frac{\sigma_{adj}}{r}\frac{\veL\veS}{2\mu^2_0}
   \label{7.5}
   \ee

   Now we come to the point of perturbative contributions to spin
   splittings.

   In heavy and  light quarkonia these contributions are simply
obtained from (102-106) substituting for $D, D_1$ their perturbative
expressions to the order $O(\alpha_s)$  (109). In case of glueballs
one should make replacement $C_2(fund)\to C_2(adj)$ in (110) and in
addition take into account two corrections due to the fact that  $i)$
valence and exchanged gluons are identical $ii)$ there is 4-gluon
vertex. These corrections have been taken into account in [64] and
 amount to some reduction of coefficients in (110).

 The resulting
matrix elements in (\ref{7.1}) look like \be
M_{SL}^{(pert)}=\frac{3C_2(adj)}{4\mu_0^2}\langle
\frac{\alpha_s}{r^3}\rangle \label{7.9} \ee \be
M_{SS}^{(pert)}=\frac{5\pi C_2(adj)}{3\mu_0^2}\langle \alpha_s\delta^{(3)}(r)\rangle
\label{7.10}
\ee
\be
M_{T}^{(pert)}=\frac{C_2(adj)}{\mu_0^2}\langle \frac{\alpha_s}{r^3}
(3\veS^{(1)}\ven\veS^{(2)}\ven -\veS^{(1)}
\veS^{(2)})\rangle
 \label{7.11}
  \ee

>From (\ref{7.10}) one can see that $M_{SS}$ can be written as
\be
M_{SS}=\frac{5\alpha_s}{4\mu^2_0}|R(0)|^2
\label{7.12}
\ee

To make simple estimates, we shall neglect first the interaction due to
perturbative gluon exchanges between valent gluons. Indeed
it was shown in [39]
 that this interaction cannot be written as Coulomb potential
between adjoint charges, and comparison to perturbative Lipatov pomeron
theory [56] shows that it is much weaker than Coulomb potential.
Neglecting this interaction altogether, one gets the lower bound of
spin-dependent effects, since all matrix clements, like
$\langle\delta^{(3)}(r)\rangle,\langle\frac{1}{r}\rangle,\langle\frac{1}{r^3}\rangle$
are enhanced by attractive Coulomb interaction.

For purely linear potential one has simple relation, not depending on radial
quantum number $n_r$ [44,47]
\be
|\Psi(0)|^2=\frac{|R(0)|^2}{4\pi}=\frac{\mu_0\langle
V'(r)}{4\pi}=\frac{\mu_0\sigma_{adj}}{4\pi}
\ee

Using (72,73),  one
obtains \be M_{ss}=\frac{5\alpha_s\sigma_{adj}}{M_0} \ee and for
$n_r=0,1$ and $\alpha_s=0.3$ one obtains
\be M_{ss}(n_r=0)=0.3 GeV,
M_{ss}(n_r=1)=0.20 GeV \ee
For $M(0^{++})$ and $M(2^{++})$ one has
the values given for the
sake of comparison with lattice calculations in Table 9 for
$\sigma_f=0.23 GeV^2$ and $\alpha_s=0.2 (0.3)$

For $L > 0$ one needs to compute spin corrections $M_{SL}$ and
$M_T$. First of all one can simlify matter using the equation (it is
derived in the same way, as (156) was derived in [47])
 \be
L(L+1)\langle\frac{1}{r^3}\rangle=\frac{\mu_0}{2}\langle V'(r)\rangle
\ee
For $V(r)=\sigma_{adj}r$ both $M_{SL}^{(pert)}$ and $M_T^{(pert)}$
are easily calculated and used in Table 9.

The estimate of $\triangle H(Thomas)=\frac{\vec S\vec
L}{2\mu_0^2}\langle\frac{V'_1+V'_2}{r}\rangle$ is more cumbersome since this
matrix element is very sensitive to the behaviour of $V_i(r)$ at small r,
where the asymptotics (151) is not yet achieved. Therefore one has to
use explicit expressions
(102, 103) of $V'_1, V'_2$ through correlators $D,D_1$
(see [50] for details and discussion).

The resulting figures for $\triangle M_{thomas}$ are used  in Table
9.  Combining all corrections and values of $M_0$ from Table 2 one
obtains glueball masses shown
and compared with lattice data in Table 9 for $\sigma_f=0.23 GeV^2$.

   The general feature of spin--dependent contribution $\Delta H_s$
   is that it dies out  fast with the growing orbital or radial
   number, which can be seen in the appearance of the $\mu^2_0$
   factor in the denominator of (\ref{7.4}).

   Indeed, from (73) one can derive that  $M_0\approx 4\mu_0$ and
   therefore   $\Delta H_s\sim \frac{1}{M^2(n,l)}\langle O(\frac1r)\rangle $, where
   $O$ stands for terms like $const. \frac1r$ or $const'. \frac{1}
   {r^3}$
   (from perturbation theory). Hence spin splittings of the radial
   recurrence of states $0^{++}, 2^{++}$ or $0^{-+}, 2^{-+}$ should
   be  smaller than the corresponding ground states.
   This feature is well supported by the lattice data in Table 9.

 We end this section with the discussion of heavy--light mesons,
 which can be calculated also in the present approach, see [46].
 While in [3] is  given   another and more exact approach for heavy--
 light mesons, taking into account chiral  symmetry breaking, here we
 shall only demonstrate that the present approach enables one not
 only reproduce the spectrum, but also calculate more delicate
 characteristics, e.g.  decay constants $f_m, M=B,B_s,D, D_s$ and
 lepton width. These  are characteristics computed in [46] and shown
 in Table 10.  One can see a reasonable agreement of all entries
 with lattice data and experiment.

  We do not touch in these lectures the topic of heavy quarkonia,
  which has been
  also extensively studied in the framework of the present
  approach [19,32], since this is the subject of lectures of
  F.J.Yndurain at this School [66].


\section{Light--cone Hamiltonian   and spectra for mesons.}

  In this chapter we follow mostly the work done in [41].
          The light--cone description of the hadron wave functions
is widely used, since it allows to get a direct connection to the parton
model and its QCD improvements \cite{feynm}. The latter however are mostly
perturbative and nonperturbative contributions are introduced
via OPE and QCD sum--rules. In this way the concept of the string -- the
main nonperturbative QCD phenomenon -- is totally lost. On physical grounds
it seems that the string is the essential ingredient of the dynamics
for large (${r\ge 0.3\>fm}$) distances, and it would be interesting
to understand its contribution to the light--cone wave functions,
formfactors, structure functions etc.

In particular, what is the QCD string in the parton language? Should one
associate it with the gluon contribution as an assembly of gluons
compressed inside the string -- or with the constituent quark mass?

There two contrasting points of view have been proposed already decades ago
\cite{kutwei,altar1}.
In \cite{kutwei} the sea quarks and gluons enter as separate entities
and one could associate the gluon distribution with the string
in the same way as photons with the Coulomb field of the charge
in the Williams-- Weizs\"acker method. In contrast to that
in \cite{altar1} the quarks have been considered as constituents with
structure, and string does not appear separately.
Recently a quantitative analysis was performed \cite{altar2}
of quark distribution in the pion starting from that in the nucleon
and assuming the same internal structure of quarks in nucleon and in pion.
It is still an open question how the structure of the constituent quarks
is formed, and to which extent it can be explained by the adjacent piece
of the string.
This problem can be elucidated partly in the present
approach, since our light--cone Hamiltonian contains the string on the light
cone explicitly.
It allows to separate the contribution of the string to the
parton distribution, in particular to the momentum sum--rule.
There is another source of structure in the constituent quark -- chiral
symmetry breaking which creates the chiral part of constituent quark.
This problem will not be discussed below, see e.g. \cite{sim1}.

Let us make a few comments about the reasons why the light-cone dynamics
is interesting
to explore in QCD. One simple reason  was stated in the beginning: it
is to formulate in terms of parton language  and  therefore to single
out the new nonperturbative contribution in the familiar parton
picture. Second reason is quite general: the necessity of using Hamiltonians
in moving frames to calculate observables, like form-factors or
cross sections, where overlap integrals of eigenfunctions in different frames
enter.
An important advantage of the light--cone wave functions is that they allow
to calculate formfactors and structure functions directly, without additional
boosts, which typically are the dynamical ones, i.e. require the use of the
exact Hamiltonian.

There is still another and deeper reason for using the light-cone
formalism. It is based on the expectation \cite{br} that the field
theory in
general has a more simple vacuum structure on the light-cone and in
addition some classes of diagrams (containing backward--in--time motion) are
suppressed. Hence one may expect that on the light-cone it is possible to
choose the field  degrees of freedom which are independent and
simplest.

We consider the relativistic quark--antiquark pair with the masses
$m_1$ and $m_2$ connected by the straight--line Nambu--Goto string with
the string tension $\sigma$ in 3+1 dimensional space--time.

Given a $q\bar{q}$ Green's function in the coordinate space
$G(x\bar{x};y\bar{y})$, where $x\bar{x}(y\bar{y})$ are final (initial)
4--coordinates of quark and antiquark, one can define the Hamiltonian $H$
through the equation (in the euclidean space--time)
\be
\frac{\partial G}{\partial T} = - HG
\ee
where $T$ is an evolution parameter corresponding to some choice of a
hypersurface $\Sigma$. In a particular case of the c.m. Hamiltonian the
role of $T$ is played by the center--of--mass euclidean time coordinate
$T=(x_4+\bar{x}_4)/2$ and the hypersurface $\sum$ is a hyperplane
$x_4=\bar{x}_4=const.$

With the notations  for the vectors $a_{\mu},b_{\mu}$
$$
ab= a_{\mu}b_{\mu} = a_ib_i-a_0b_0=a_{\bot}b_{\bot}+a_+b_-+a_-b_+,
$$
$$a_{\pm}=\frac{a_3\pm a_0}{\sqrt{2}},$$
one can define the hypersurface $\sum$ through the $q\bar{q}$ coordinates
$z_{\mu}.\bar{z}_{\mu}$ as
$$
z_+(\tau)=\bar{z}_+(\bar{\tau})
$$
and the kinetic part of the action $A$
$$
A=K+\bar{K}+\sigma S_{min},
$$
has the form
$$
K+\bar{K} = \frac{1}{4} \int^s_0 \dot{z}^2_{\mu}(\tau)d\tau+
\frac{1}{4}
 \int^{\bar{s}}_0 \dot{\bar{z}}^2_{\mu}(\bar{\tau})d\bar{\tau}+
  \int^{s}_0m_1^2 d\tau+ \int^{\bar{s}}_0m_2^2 d\tau =
$$
$$=
\int^T_0dz_+\left[ \frac{\mu_1}{2}(\dot{z}^2_{\bot}+2\dot{z}_-)+
\frac{\mu_2}{2}
(\dot{\bar{z}}^2_{\bot}+2\dot{\bar{z}}_-)
+\frac{m_1^2}{2\mu_1}+\frac{m^2_2}{2\mu_2}\right]
$$
  where we have defined
$$
  2\mu_1(z_+)=\frac{\partial z_+}{\partial \tau}~;~~
  2\mu_2(z_+)=\frac{\partial \bar{z}_+}{\partial \tau}
$$

The minimal surface $S_{min}$ is formed by connecting $z_{\mu}(z_+)$ and
$\bar{z}_{\mu}(z_+)$ with the same value of the evolution parameter $z_+$, i.e.
$$
  S_{min}= \sigma \int^T_0 dz_+\int^1_0 d\beta [\dot{w}^2
  w'^2-(\dot{w}w')^2]^{1/2}
$$
  where
$$
  w_{\mu}(z_+; \beta) = z_{\mu}(z_+) \beta + \bar{z}_{\mu}(z_+) (1-\beta)
$$
  and dot and prime denote derivatives  in  $z_+$ and $\beta$ respectively.

  We now introduce "center--of--mass" and relative coordinates,
$$
  \dot{R}_{\mu}=x\dot{z}_{\mu} + (1-x)\dot{\bar{z}}_{\mu}~,~~
  \dot{r}_{\mu}=\dot{z}_{\mu}-\dot{\bar{z}}_{\mu}
$$
where the variable $x$ is defined
 from the requirement that the term
$\dot{r}_{\bot}\dot{R}_{\bot}$ should be absent in the action.
This yields:
\be
x=\frac{\mu_1+\int \nu\beta d\beta}{\mu_1 + \mu_2 + \int \nu d\beta}
\label{prex}
\ee
Then for the Green function one obtains:
$$
G(x\bar{x};y\bar{y})= \int D{\mu}_1(z_+) \; D\mu_2(z_+)D\nu \;
DR_{\mu} \; Dr_{\mu} \; e^{-A}
$$
where the action $A$
\begin{eqnarray}
A=\frac{1}{2}\int dz_+\left\{\frac{m^2_1}{\mu_1}+\frac{m^2_2}{\mu_2}+
a_1(\dot{R}^2_{\bot}+2\dot{R}_-)+a\dot{r}^2_{\bot}+ \right.
\nonumber \\
+ \int\frac{\sigma^2}{\nu}d\beta \; r^2_{\bot} -
\frac{(r_-+\dot{R}_{\bot}r_{\bot}+(\langle \beta\rangle -x)\dot{r}_{\bot}r_{\bot}
)^2}{r^2_{\bot}(\int \nu d\beta)^{-1}}-
\nonumber \\
\left. - \frac{(\dot{r}_{\bot}r_{\bot})^2\int\nu(\beta-\langle \beta\rangle )^2d\beta}
{r^2_{\bot}}
\right\},
\label{act}
\end{eqnarray}

The following notation was used:
\be
a_1=\mu_1+\mu_2+\int^1_0\nu(\beta)d\beta
\ee

\be
a=\mu_1(1-x)^2+\mu_2x^2+\int^1_0 \nu(\beta)(\beta-x)^2 d\beta
\label{adef}
\ee
Integration over $DR_{\mu}$ leads to an important constraint:
$$
a_1 = P_+
$$
Furthermore we go over into the minkowskian space, which means that
$$\mu_i\to - i\mu^M_i~,~~\nu \to - i\nu^M~
, ~~A\to - i A^M
 $$
 For the minkowskian action we obtain  (omitting from now on the superscript
 $M$ everywhere)
 $$
 A^M= \frac{1}{2} \int dz_+\left\{
 -\frac{m^2_1}{\mu_1}-\frac{m^2_2}{\mu_2}+a\dot{r}^2_{\bot}- \int
 \frac{\sigma^2 d\beta }{\nu}r^2_{\bot}-\right.
$$
$$
  -\left.\nu_2\frac{(\dot{r}_{\bot}r_{\bot})^2}{r^2_{\bot}}-
 \frac{\nu_0a_1}{(\mu_1+\mu_2)r^2_{\bot}}[r_-
+(\langle \beta\rangle -x)\dot{r}_{\bot}r_{\bot}]^2 \right\}
 $$

To complete the Hamiltonian formulation of our problem we define canonical
momenta for the coordinates $\dot{R}_-$ and $\dot{r}_-$.
As it was shown, canonically conjugated momentum to the
$\dot{R}_-$ is $P_+=a_1$. Situation with $\dot{r}_-$ is more subtle.
In order to clarify the situation let us start with the general
form of the $q\bar{q}$ Green's function in the
Feynman--Schwinger formalism
$$
G(x,y)=\int ds~d\bar{s}DzD\bar{z}`^{-K-\bar{K}}\langle W(C)\rangle
$$
to impose boundary conditions, one can rewrite $DzD\bar{z}$ using
discretization
$$ \xi_n\equiv z(n)-z(n-1), \bar{\xi}_n=\bar{z}(n)-\bar{z}(n-1)$$
$$
DzD\bar{z} = \prod_{n,n'}d\xi_nd\bar{\xi}_{n'} \;  dp~dp' \times
$$
$$
\times exp \left\{
ip(\sum\xi_n + X-Y) + ip'(\sum\bar{\xi}_{n'}+X-Y) \right\}
$$
One can introduce the total and relative momenta
$$
P=p+p',~~q=\frac{p-p'}{2};
$$
and $\dot{R}=\Delta R_n /\varepsilon; ~~N\varepsilon
=T, $  one has
 $$
 \xi_nx_n+\bar{\xi}_n(1-x_n)=\Delta R_n
 $$
$$
 \xi_n-\bar{\xi}_n=\Delta r_n
 $$
Expressing $\xi_n,\bar{\xi}_n$ through $\Delta R_n, \Delta r_n$
 and going over to the momentum representation of $G$ one obtains
 $$
 G(P)=\int dq \prod_{n,n'} d\Delta R_n d\Delta r_{n'} \times
$$
$$
 \times exp \left\{ iA + iP\sum_{n}\Delta
 R_n+i\sum_n(\frac{1}{2}P(1-2x)+q)\Delta r_n \right\}
$$
 where $A=\int^T_0d\tau{\cal L} = \int^T_0 d\tau (K+\bar{K}-\sigma
 S_{min})$
 One can now introduce the Hamiltonian form of the path integral via
 $$
 \int Dx e^{i\int {\cal L}d\tau }= \int Dx Dpe^{ip_k\dot{x}_k-i\int {\cal
 H}d\tau}
 $$
 and rewrite the first two exponents as
 $$
 exp \left\{ i\int P_i \dot{R}_i d\tau + i \int[\frac{1}{2} P_i
 (1-2x(\tau))+q_i]\dot{r}_id\tau \right\}
$$
 The term proportional $q_i$ disappears because of boundary conditions
 $r_{\mu}(0) = r_{\mu}(T) =0$, and one obtains
 $$
 p_+=\frac{1}{2} P_+(1-2x),~~[p_+,r_-]=-i
 $$
 One can rewrite this in the form
 $$
 P_+r_-=i\frac{\partial}{\partial x}, ~~[P_+r_-, x]=i
 $$

 Separating out the center of mass motion one defines the Hamiltonian
from the
 corresponding Minkowskian action $A^M$:
\be
A^M =  \int dz_+ L^M~, ~~H=p_{\bot}\dot{r}_{\bot}-L^M
\label{mink}
\ee
with $p_{\bot} = \partial L^M / \partial \dot{r}_{\bot}$.

After all one easily obtains the explicit form of the Hamiltonian
\begin{eqnarray}
 H=\frac{1}{2} \left\{ \frac{m^2_1}{\mu_1}+\frac{m^2_2}{\mu_2}+
 \frac{L_z^2}{a\>r^2_{\bot}}+
 \frac{(p_{\bot}r_{\bot}+\gamma r_-)^2}{\tilde{\mu} r^2_{\bot}}\right.
\nonumber \\
+\left.\int \frac{\sigma^2}{\nu} d\beta r^2_{\bot}+ \frac{\nu_0 P_+}{\mu_1+\mu_2}
 \frac{r^2_-}{r^2_{\bot}}  \right\}
\label{ham}
\end{eqnarray}

Let us remind, that
$\mu_1, \mu_2 $ are einbein fields, playing the role of $P_{+}$
momenta of the particles, $\tilde{\mu} =
\mu_1 \mu_2 / (\mu_1 + \mu_2)$
,  $\nu_0 = \int_{0}^{1} d\beta \nu(\beta)$ and $\nu(\beta)$ is
the einbein field with the physical meaning of the $P_{+}$--momentum,
carried by the string.
As it has been mentioned above, the Hamiltonian explicitly depends
on $L_z^2 = {\vec p}~^{2}_{\bot}{\vec r}~^{2}_{\bot} -
({\vec p}_{\bot}{\vec r}_{\bot})^2
$, the corresponding mass parameter is equal to (\ref{adef}):
$$
a=\mu_1(1-x)^2 + \mu_2 x^2 + \int\limits_0^1 d\beta \nu(\beta)
(\beta - x)^2
$$
The variable $x$ in the above expression is the same as defined in
(\ref{prex}):
 \be x=\frac{\mu_1 + \langle \beta\rangle \nu_0}{P_+}
 \label{x}
 \ee
 where
 \be
 P_+ = \mu_1 + \mu_2 + \nu_0
\label{P}
\ee
is the light--cone total momentum of the system.
Also
\be
\gamma = \nu_0\left(\langle \beta\rangle  -
\frac{\mu_1}{\mu_1 + \mu_2}\right)
\label{gamm}
\ee
where $\langle \beta\rangle  = \int_0^1 d \beta \beta \nu(\beta) / \int_0^1 d \beta
\nu(\beta)$. It is easy to show, that in the limit ${\sigma} \ll {m^2}$
Hamiltonian turns out to have well known form:
\be
H = {\frac{1}{2P_+}}\left( \frac{m_1^2}{x} + \frac{m_2^2}{1-x}\right)
 + {\sigma}\cdot |r|
\ee
Let us stress that separation of this kind
 cannot be done beyond nonrelativistic
limit.

As the next step one should quantize the classical Hamiltonian function.
Before doing it,
one should choose the appropriate set of dynamical variables.
Three einbein fields ${\mu}_1 , {\mu}_2 , \nu $ introduced above play
different dynamical roles. In the
nonrelativistic case $m_1, m_2 \gg \sqrt{\sigma}$ (and therefore for the free
particles) the dependence of Hamiltonian (and wave functions) on $\nu$ can be
correctly found by minimization procedure with $\nu$ taken as a classical
variable.  This is in its turn a consequence of the fact, that string in our
approach is the minimal one -- it has no internal degrees of freedom and may
only stretch or rotate as a whole.

On another hand, $\mu_1$ and $\mu_2$ on the light cone play the role
of legitimate quantum dynamical degrees of freedom and can be expressed
through $x$ and $P_+$ as in (\ref{P}) and (\ref{x}).

There are two canonically conjugated pairs in our problem:
$\{{\vec p}_{\bot}, {\vec r}_{\bot}\}$ and $\{x, (P_+r_-)\}$.
We introduce also a new
dimensionless variable $\tilde y$ instead
of $\nu$ : $\tilde y (\beta) = {\nu}(\beta)/{P_+}$.
It satisfies the obvious condition:
$0<\tilde y<1$. This variable depends on ${\vec r}_{\bot}$ as well as on
$(P_+r_-)$.
Rigorously speaking one should extract this dependence by the minimization
of the Hamiltonian with respect to $\tilde y$ as it has been explained above.
Instead an easier (however approximate) way is chosen. On physical
grounds it can be shown that for small
$r^2_{\bot}$ one has ${\tilde y}_0 = \int_0^1 \tilde y d\beta \sim
const\cdot r^2_{\bot}$ so that linear string energy density
${\tilde y}_0/r^2_{\bot}$ stays finite if $r_{\bot}^{2} \to 0$,
and using this property one can
reproduce the correct 1+1 limit, namely 't Hooft equation \cite{thof},
from the 3+1 light--cone Hamiltonian (\ref{ham}) (see \cite{dub1}.
On the other hand, if distances are increasing, ${\tilde y}_0$ tends to some
limiting value which is determined by the virial theorem arguments.
So one can parameterize ${\tilde y}$ introducing several parameters
and replace the minimization in the functional sense by the ordinary
minimization with respect to these parameters.
We have chosen the simplest 2--parametric form
for ${\tilde y}$:
\be
\tilde y = \frac{yt}{1+\alpha t}
\label{y}
\ee
where $t={r^2_{\bot}}$ and $y$ and $\alpha$ are free parameters.
The requirement $0 < \tilde y < 1$ leads to the restriction
$0 < y < \alpha $. Let us stress again that this parameterization is
the matter of convenience and all physical results are determined
from the requirement that every energy level should have its own minimum.

It is easy to see from (\ref{x}), that the $x$--variable  is
the part of the total momentum, carried by the first quark itself and a
part of the string, "belonging" to this quark.
Rewriting (\ref{x}) in the form:
\be
\frac{\mu_1}{P_+} = x - \tilde y \langle \beta\rangle  \;\;;\;\;\frac{\mu_2}{P_+} = (1-x) -
\tilde y (1-\langle \beta\rangle )
\label{muP}
\ee
one can conclude, that for the given $\tilde y$ the variable $x$ may vary in
the following limits:
\be
\tilde y \langle \beta\rangle \; \le\; x\; \le\; 1 - \tilde y \>(1 - \langle \beta\rangle )
\label{yt}
\ee
It is more convenient to make rescaling to a new variable $\rho $ which
varies from zero to unity:
\be
x = \tilde y \langle \beta\rangle  + (1-\tilde y) \rho\;\; ; \;\; 0\le\rho\le 1
\label{r}
\label{xt}
\ee
The quantity we are really interested in is the mass operator squared:
\be
{\hat M}^2 = 2\>{\hat H}\>P_+
\label{msq}
\ee

Using definition (\ref{x}) we obtain the following form for $M^2$:
$$
 M^2= \left\{ \frac{m^2_1}{x-{\tilde y}_0 \langle \beta\rangle }+
\frac{m^2_2}{1-x - {\tilde y}_0 (1-\langle \beta\rangle )}+
 \frac{L_z^2}{{\tilde c}\>r^2_{\bot}} + \right.
$$
$$
 + \left[ \frac{1}{x-{\tilde y}_0 \langle \beta\rangle } + \frac{1}{1-x - {\tilde y}_0
(1-\langle \beta\rangle )} \right] \times
$$
$$
\times \frac{[p_{\bot}r_{\bot}+ \tilde {\gamma} (P_+r_-)]^2}{ r^2_{\bot}} +
$$
$$
+ \left.\int \frac{\sigma^2}{\tilde y} r^2_{\bot} d\beta +
\frac{{\tilde y}_0}{1-{\tilde y}_0} \frac{(P_+r_-)^2}{r^2_{\bot}}  \right\}
$$
here
$$
{\tilde c} =
(x-{\tilde y}_0\langle \beta\rangle )(1-x)^2 +
$$
$$
[1-x -{\tilde y}_0 (1-\langle \beta\rangle )] x^2 + \int_0^1\tilde y (\beta - x)^2 d\beta
$$
and
$$
\tilde \gamma = \frac{{\tilde y}_0}{1-{\tilde y}_0}\> (\langle \beta\rangle  - x)
$$
As it was mentioned above, it is more convenient to express $M^2$ through
a new variable ${\rho}$:
$$
\rho = \frac{1}{
1 - {\tilde y}_0} (x - {\tilde y}_0 \langle \beta\rangle )
$$
In terms of this variable one obtains:
\begin{eqnarray}
 M^2= \frac{1}{1-{\tilde y}_0} \left\{ \frac{m^2_1}{\rho}+
\frac{m^2_2}{1-\rho}+
\frac{L_z^2}{{\tilde c}\>t} + \right.
\nonumber \\
+ \left( \frac{1}{\rho} + \frac{1}{1-\rho} \right) \times
\frac{[p_{\bot}r_{\bot}+ \tilde {\gamma} (P_+ r_-)]^2}{t} +
\nonumber \\
+ \left.(1-{\tilde y}_0) \int \frac{\sigma^2}{\tilde y} t d\beta +
\frac{{\tilde y}_0 (P_+r_-)^2}{t}  \right\}
\label{mm}
\end{eqnarray}
where the notation $t = r_{\bot}^2$ was used.

Quantization of the above expression  is done according to the canonical
commutation relations:
$$
\{p_{\bot}^{k}, r_{\bot}^{j}\} = -i {\delta}^{kj}
$$
$$
\{x, (P_+r_-)\} = -i
$$
We are looking for the wave function of the problem given in the mixed
coordinate -- momentum representation $\psi = \psi(\rho, t)$, so one has
to substitute into (\ref{mm}) the operators:
$$
(P_+r_-) = i\> \left(\frac{1}{ 1-{\tilde y}_0}\right)
\frac{\partial }{ \partial \rho}\;
\; ;\;\; p_{\bot}^k = -i\>\frac{\partial }{ \partial r_{\bot}^k}
$$
The important point is
the operator ordering. We use the Weil ordering rule, i.e.
$$
AB \to \frac12 (\hat A \hat B + \hat B \hat A)
$$
for any noncommuting operators $A$ and $B$.
Let us notice that $\tilde y$ explicitly depends on $t$ according
to (\ref{y}) and hence should also be differentiated during the ordering
procedure.
The final result for the operator ${\hat M}^2$ may be found by the
straightforward calculations, it is
\begin{eqnarray}
{\hat M}^2 =
A_1\>\frac{{\partial}^2}{{\partial t}^2} + A_2 \>
\frac{{\partial}^2}{{\partial \rho}^2} + A_3 \>
\frac{{\partial}^2}{{\partial t}{\partial \rho}} + \nonumber \\
+ A_4\>\frac{\partial}{\partial t}  + A_5 \frac{\partial}{\partial \rho}
 + A_6
\end{eqnarray}
where  $A_i$ are expressed through $\rho, t,y$, (see \cite{dub1} for
more details).

We have chosen six sets of quark mass parameters, including 4 sets
of equal masses and 2 sets of unequal masses in the interval from 0
to 5 GeV.  The values of parameters $y, \alpha, \epsilon$ are
obtained by minimization procedure as explained above.  The
eigenstates obtained by our numerical procedure are
 shown in Fig.2.

Let us first discuss the light--cone spectrum. The Hamiltonian (\ref{ham})
depends explicitly on $L_z$, whereas the Lorentz--invariance requires
that masses do not depend on $L_z$, but depend on $L$. Consequently one
expects
that the lowest state for $L_z = 0$ corresponds to $L=0$, while the next
state with $L_z = 0$ corresponds to $L=1$ and must be degenerate in mass
with the state $L_z = 1 ,  L=1$. This is clearly seen in
the data.

At this point one should take into account another degeneration -- the dynamical
one, and to this end discuss first center--of--mass spectrum, obtained from the
c.m. Hamiltonian
 \cite{feynm}.
 The c.m. Hamiltonian for $L=0$ reduces to
the so--called spinless Salpeter equation \cite{br} which was
actually solved by us. The approximate form of the spectrum
 can be
represented as:
 \be \left( M^{(0)}(L, N_r)\right)^2 \cong 2\pi
\sigma \left( 2N_r + \frac{4}{\pi} L + \frac32 \right) \label{nhn}
\ee
For  $L>0$ one should take into account the string contribution, absent
in spinless Salpeter equation but present in the Hamiltonian
\cite{feynm}. This correction
 is done in (113).
 The corrected masses,
$M(L, N_r) = M^{(0)} + \Delta M $ are shown in Fig.2.
  Now one can
see that there is an approximate degeneration in mass $M = M^{(0)} +
\Delta M$ of states when one replaces one unit of $N_r$ by two units
of $L$.
The same type of approximate mass degeneration is seen
in the light--cone spectrum -- compare e.g. the states with $(N_r ,
L, L_z) = (1,0,0)\; ; \; M_{LC} = 2.25$ and $ (0,2,0) \; ;\; M_{LC} =
2.28$.

In Fig.2 this degeneration is visualized as the fact that
masses appear on the vertical lines. This degeneration is a dynamical
one and is a characteristic feature of nonrelativistic oscillator. In
our relativistic case it reveals a new string--like symmetry,
typical for the QCD string spectrum \cite{sim2}.

Let us now compare the light--cone and the center--of--mass spectrum.
One could expect the coincidence up to an overall shift due to the
different treatment of Z--graphs in two systems. In the c.m. Hamiltonian
these Z--graphs are presented but supposed to be unimportant on the
grounds, that
the backtracking of a quark in time necessarily brings about a folding in
the string world sheet which costs a large amount of action and is
therefore suppressed. The situation is different in the light cone --
it is general belief that Z--graphs are absent here.
One also expects that the Z--graphs (and the overall shift)
should decrease if quark masses are increasing.

The comparison of the spectra can be made from Fig. 2. One
can see indeed some overall shift down in the c.m. spectrum by some
0.1 GeV and otherwise the masses coincide within the accuracy of
computation.  This fact is highly nontrivial since two quantum
Hamiltonians (the light--cone and center--of--mass ones) are
different, they cannot be obtained from each other by a simple boost
or other simple transformation.  The light--cone Hamiltonian is
rather complicated and it took the authors more than a year to get
reasonable numerical results for it.

We have also checked the quark mass dependence of the overall shift of
spectra and proved that it drops sharply with the quark mass increasing
supporting the idea, that the shift is due to different treatment of
Z graphs (or self energy graphs). This fact also confirms, that the shift
is not a consequence of some systematic errors of our procedure.

We now turn to eigenfunctions. One expects in this case two types of excitation:
the radial one leading to new nodes on $\rho$ coordinate and similar to
the 1+1 excited states and the $r_{\bot}$--excitation which causes  nodes in
the $t$--coordinate and associated with the orbital excitation.
This is clearly seen in Fig. 3, where (a) refers to the ground
state, (b) --
to the orbital and (c) -- to the radial excitation. A more complicated example,
combining both types of excitation is given in Fig. 3 (d).

As the next illustration we show in Fig. 4 the case of two heavy
quarks,
demonstrating two types of excitation and also a new feature --
the actual region of parameters is squeezed to a small region near
$\rho = 1/2$.

At this point it is important to find connection of our light--cone wave
function to the nonrelativistic one, usually defined in the c.m..
In [5] it was demonstrated that this connection can be established
only when both quarks are heavy, $m_1, m_2 \gg \sqrt{\sigma} $.
In this case $\tilde y \ll 1$ (as can be found by direct minimization
of the Hamiltonian (\ref{ham})) -- the string transforms into the potential
and loses its material and momentum contents. One can introduce as in
[5] the relative momentum $p_z$ and relative coordinate $r_z$:
\be
p_z = (m_1 +m_2) \left( x - \frac{m_1}{m_1 + m_2} \right)
\label{pz}
\ee
\be
r_z = \frac{P_+ r_-}{m_1 + m_2}
\label{io}
\ee
and the $M$ operator can be written as:
\be
 M \approx m_1 + m_2 + \frac{1}{2 \tilde{m}}
({\vec{p}}~^{2}_{\bot} + {\vec{p}}~^{2}_{z}) +
\sigma \sqrt{r^{2}_{\bot} + r^{2}_{z}}
\label{Mapp}
\ee

Hence one can write the momentum--space nonrelativistic wave function
$\Psi({\vec p}~^{2})$ directly through the light--cone variables:
\be
\Psi({\vec{p}}~^{2}) = \Psi \left[p_{\bot}^2 + (m_1 + m_2)^2\> \left(
 x - \frac{m_1}{m_1 + m_2} \right)^2 \right]
\label{psip2}
\ee

This representation is valid in the large mass limit $m_i \gg \sqrt{\sigma}$
stated above and in addition near the center of the $x$--distribution,
i.e. when $|x - m_1 / (m_1 + m_2)| \ll 1$. The width of the peak in $x$
variable is proportional to $(m_1 + m_2)^{-2}$ and is very narrow for
heavy quarks. The form (\ref{pz}) is not correct for $x$ at the ends of the
interval, i.e. $x = 0,1$ (remember that for large $m_i$ the extremum value
of $\tilde{y}$ tends to zero and $x = \rho$ changes in the interval
[0,1]). Indeed the exact wave function as discussed in Section~4 vanishes
linearly at $x = 0,1$, while the r.h.s of (\ref{psip2}) stays nonzero.
Moreover, the Jacobean {$\cal{J}$} of the phase space
$d^3 p = {\cal{J}} dx d^2 p_{\bot}$ is constant ${\cal{J}} = m_1 + m_2$
and does not change this conclusion.

The correspondence of the c.m. and light--cone wave functions is lost
if quark masses $m_i$ are of the order of $\sqrt{\sigma}$ or less.
The physical reason is that the role of dynamics is now 100\%
important and the dynamics is {\underline {different}} in different
frames: the light--cone Hamiltonian and wave functions are connected
with those of the center--of--mass by a dynamical transformation,
which includes {\underline {nonkinematical}} Poisson operators.
Hence one cannot hope to obtain one wave function from another
using only transformation including the kinematical (free) part of the
operator $M^2$. Moreover, a glance at the operator $M^2$ in (\ref{ham})
helps to realize that the separation of $M^2$ into the purely
kinetic part (containing only momentum and masses) and purely
potential one (containing only relative coordinates) is
{\underline {impossible}} at all: e.g. the string enters through
the term $(P_+r_-)^2$ and mixed terms like $p_{\bot}r_{\bot}$
are present everywhere. This circumstance limits the
use of simple recipes known in literature \cite{chung},
which connect the c.m. and light--cone wave functions.
In particular, the ansatz for $p_z$ suggested in \cite{chung}
and used for equal quark masses
\be
p_z = \sqrt{\frac{m^2 + k_{\bot}^2}{x(1-x)}} \left( x - \frac12 \right)
\label{g}
\ee
coincides with our form (\ref{io}) rigorously derived in \cite{dub}
only for large $m$, $m \gg \sqrt{\sigma} $ and $x$ in the narrow region
around $x = 1/2$ but differs at the tails of the wave function.
For light quark masses, $m \le \sigma$ and small $p_{\bot}$
the relation (\ref{g}) yields incorrect
results which can be seen in the almost constant behaviour in
the interval [0,1] of the light--cone wave function, produced by the
insertion of (\ref{g}) into the c.m. wave function, typically
$$
\psi({\vec p}~^{2}) \sim \exp (- a^2 {\vec p}~^{2})
\to \exp (- a^2 [ p_{\bot}^2 + \frac{m^2 + p_{\bot}^2}{x(1-x)}
(x-\frac12)^2 ] )
$$
is insensitive to $x$ for $m,p_{\bot} \to 0$, whereas
the exact light--cone wave function is decreasing as $x(1-x)$, see
Fig. 5.

At the same time for heavy quark masses, $m \gg \sqrt{\sigma}$ the two
functions, one transformed by (\ref{g}) from the c.m. and another is
a genuine light--cone wave--function solution of (177), are very
close to each other.

We now turn to the formfactor of computed states.
 The main feature
is a very slow decrease of $F(q^2)$ with $q^2$
which signals in particular a small radius of states. Indeed the
values of $\sqrt{\langle r^2\rangle }$ are too low ($\sim 0.338$ $fm$
for massless quarks). This fact is in agreement with the earlier c.m.
calculations of \cite{carl}.

Finally we turn to the quark--distribution function $q(\rho)$.
It is computed through the light--cone wave--function $\psi(\rho ,t)$
using  the relation $q(\rho)=\pi\int^\infty_0|\Psi
(\rho,t)|^2 dt$ and shown in Fig.  6.  One can see the
symmetric behaviour of $q(\rho)$ with respect to reflection $\rho \to
1-\rho$.  At the ends of the interval $q(\rho)$ vanishes like
$\rho^{2}$ and $(1-\rho)^2$, in the agreement with the $1/q^3$
behaviour of the formfactor at large $q$ due to the Drell-Yan-West
relations \cite{west}.

Note the narrowing of the peak in $q(\rho)$ in Fig.~6
 for
increasing quark masses.

The main physical idea of our approach is that the most part of
nonperturbative dynamics in QCD is due to the QCD string, and the latter
is described by the Nambu-Goto part of the Hamiltonian, which was written
before in the c.m. [5] as well as in the light-cone
coordinates \cite{dub}.

Only valence part of the Fock's column was considered above in the paper,
also for simplicity spins and perturbative gluon exchanges are neglected.
To do the systematic comparison with experiment all these three
simplifications should be eliminated. Let us discuss their effect point
by point. The higher Fock states are necessary to reproduce the Regge
behaviour of $q(x) \sim x^{-\alpha_{\rho}(0)}$
at small $x$ (and at $x \to 1$ for quark distributions of hadrons in
high-energy scattering). Here comes the first crucial point; to be answered
in the second part of these lectures, what is
the QCD reggeon?

In our method the higher Fock states, constituting the QCD reggeon,
correspond to several gluons propagating in the nonperturbative
background and therefore confined to the excited Nambu-Goto surface
[2].  These states are in one-to-one correspondence with
the excited Nambu-Goto string states. This is the picture at large
distances; at small distances smaller than the vacuum correlation
length (width of the Nambu-Goto string) $T_{g} \sim 0.2$ $fm$,
 the string disappears and the usual perturbative gluon
exchanges reappear.

The effect of spin of light quarks is highly nontrivial
[3].  It leads to the creation of the new vertex, which
yields the constituent quark structure. Physically one may imagine
this structure as being due to the light--quark walks around the end
of the string.

Having said all this, what are the lessons of the present work and of it
possible development?

The first lesson is that valence quark component
can be successfully dynamically computed on the light cone; the
minimal string on
the light cone is physically and mathematically well defined. The
structure of the spectrum
obtained on the light cone for the first time reasonably coincides with that
of the c.m. Hamiltonian for the string with quarks.
Effects of Z-graphs are estimated to be less than 12\% even for light quarks.
Moreover, the nonperturbative
wave function obtained directly on the light cone allows to calculate
nonperturbative contributions to the formfactor and structure function.

The second lesson is that the formfactor computed directly on the light cone
is close to of c.m. for small $q$, but is systematically
above the c.m. formfactor for larger $q$. This is not surprising, since in the
light--cone formfactor there is a mechanism of the "redistribution" of the
momentum $q$ between the quarks, since $q$ enters the light--cone formfactor
 multiplied with $(1 - x)$, so that the larger $q$, the smaller
is $(1 - x)$ and the wave function does not decrease too fast.
Physically it means that at large $q$ the configuration survives where
the spectator quark gets as little momentum as possible so that it can
be easily turned together with the active quark. This is exactly what
is called the Feynman mechanism [76].

The third lesson is that the minimal string plays only a passive role on
the light cone, namely it participates in sharing of the total momentum
and carries the part equal to $\langle \tilde{y}\rangle $, but it does not produce the
$x$--distribution in structure function,
which could simulate the gluonic structure
function. The reason is that the string variable -- the einbein field
$\nu$ -- is quasiclassical and has no dispersion.

The value of $\langle \tilde{y}\rangle $ computed according to
relation $<\tilde> =\pi \int^\infty_0\int^1_0 \tilde|\Psi (\rho,
t)|^2 d\rho dt$ depends on quark masses and is equal to 0.22 for
massless  quarks.  It is resonable that
$\langle \tilde{y}\rangle = 0.22$ is smaller than the experimental
value of overall gluon momentum, $0.55$, since in our picture the
difference should be filled in by higher Fock components.

The fourth lesson comes from the comparison of the computed quark
distribution,  with the experimental data for the pionic
structure function \cite{sutt}. Behaviour of $q(\rho)$ at small $\rho$ and
small $(1- \rho)$ is symmetric in Fig.6, while in reality
$q(x)$ should
rise at small $x$ like $x^{-\alpha_{\rho}(0)} \sim x^{-0.5}$
(we neglect at this point the difference
between $x$ and $\rho$, which is due to $\langle \tilde{y}\rangle $). This peak at the small
$x$ should be filled in by the contribution of higher Fock components,
containing additional gluons on the string, as was discussed above. The
behaviour of $q(\rho)$ at $\rho = 1$, which is calculated to be
$(1-\rho)^{2}$ will be also changed into $(1 - \rho)$ due to gluon
exchanges, which account for the formfactor asymptotic $1/q^{2}$ at large
$q$, and the Drell-Yan-West duality ensure the $(1 - \rho)$ behaviour
around $\rho = 1$.

Finally, the formfactor calculated above in the paper,  shows too
little radius of the "pion" $\langle r^2\rangle  \approx (0.34 \; fm)^2$ as compared
with the experimental one $\langle r^2\rangle  \simeq (0.67 \; fm)^2$. This fact is
in qualitative agreement with other calculations, where the
c.m. wave--function was used \cite{carl}, and some authors assumed
as in \cite{altar2} that quarks have "internal" structure and their
own radius which should be added to the "body radius" to reproduce
the experimental value. This fact of small body radius seems to
be a necessary consequence of the simple string + point-like quarks
picture, and probably cannot be cured by the higher Fock
components;in the second part of lectures we shall come to this point
fromanother direction.


\section{Conclusions}

As one can see from previous chapters, the simple relativistic string
Hamiltonian (111-112) and its extension to the hybrid and glueball
case works surprisingly well and describes quantitatively the
observed meson spectrum and calculated on the lattice spectrum of
hybrids and glueballs. This implies that confinement in the form of
the thin string is the basic element of the strong interaction, and
for the most hadrons this is enough to reproduce main features like
masses, decay constants, magnetic moments etc.
Last years there appeared new successful applications of the method
to the polarizability of mesons [78], to the description of nucleons
[79] and to the calculation of the nonperturbative correlation masses
in $SU(2)$-- Higgs model [80]. Analysis of heavy quarkonia, sensitive
to small distances adds to this picture two elements: one needs to
know one--loop corrections to spin--dependent forces and small
distance behaviour of NP forces [19], [30]. However the world of
Nambu--Goldstone bosons and chiral symmetry breaking (CSB) phenomena
was completely omitted in the previous material for the lack of
space. A new approach based on the analysis of the heavy light system
[3], allows to formulate dynamical equations for mesons and baryons
[65] in this case, taking into account CSB.
Simultaneously one can calculate in the method the NP self-energy
of the quark (constant $C_0$ subtracted from the mass eigenvalue
in (112)), and consider spin effects not assuming its smallness.
 This
part of material will be published separately as a second part of
lectures.

The author is grateful to A.B.Kaidalov, Yu.S.Kalashnikova,
F.J.Yndurain for discussions and to the Organizing Committee and
Professor Lidia Ferreira for the effective organization and warm
hospitality at the XVII International School.

 \newpage

{\bf Table 1}\\

{Comparison of the WKB spectrum of Hamiltonian (72)
$M^{(1)}_n$ with the exact spectrum of Hamiltonian (71)
$M^{(2)}_n$ for $m=0$ and $\sigma=0.2~GeV^2$.}

\begin{center}

\vspace{1cm}

\hspace*{3cm}
\begin{tabular}{|c|c|c|c|c|c|c|}
\hline
$n$&0&1&2&3&4&5\\
\hline
$M_n^{(1)}$& 1.475& 2.254& 2.825& 3.299& 3.713& 4.085\\
\hline
$M_n^{(2)}$& 1.412& 2.106& 2.634& 3.073& 3.457& 3.803\\
\hline
\end{tabular}
\end{center}

\vspace{1cm}

{\bf Table 2}\\

Effective mass eigenvalues $\mu_0(n,l)$ (in GeV for $\sigma_f=0.18 GeV^2$)
  obtained from Eq.(
  70),
  (72) $\mu_0=\sqrt{\sigma_{f}}(\frac{a(n)}{3})^{3/4}$-
  upper entry,and eigenvalues of reduced equation a(n)--lower entry.

  \vspace{1cm}

\begin{center}

\begin{tabular}{|ll|l|l|l|l|} \hline
&n&0&1&2&3 \\
L&&&&&\\\hline
0&&0.352&0.535&0.67 &0.78 \\
&&2.3381&4.0879&5.520&6.786\\ \hline
1&&0.462&0.611&0.732 &\\
&&3.3613&4.8845& 6.216& \\  \hline
2&&0.55 &0.68 & & \\
&&4.248&5.63& & \\
3&&0.627&&&\\
&&5.053&&&\\ \hline
  \end{tabular}
 \end{center}

  \vspace{1cm}

{\bf Table 3}\\

 Mass eigenvalues of the rotating string Hamiltonian (67)

  \vspace{1cm}

\begin{center}

\begin{table}[ht]
\hspace*{3cm}
\begin{tabular}{|c|c|c|c|c|c|}
\hline
\hspace*{0.3cm}$n$  $l$&1&2&3&4&5\\
\hline
0&1.865&2.200&2.481&2.729&2.956\\
\hline
1&2.562&2.832&3.068&3.281&3.480\\
\hline
2&3.091&3.329&3.540&3.733&3.913\\
\hline
3&3.535&3.753&3.947&4.125&4.290\\
\hline
4&3.925&4.128&4.309&4.476&4.629\\
\hline
5&4.278&4.469&4.638&4.797&4.939\\
\hline
\end{tabular}
 \end{table}
\end{center}

\newpage

{\bf Table 4}\\

Energy eigenvalue $a(\lambda)$ of the reduced equation (87) as
a function of the dimensionless paramenter $\lambda$, (88);
derivative $|a'(\lambda)|=\mid
\frac{da(\lambda)}{d\lambda} \mid$ and wave function
at origin $\chi_{\lambda}(0)$ (solution of (87)) divided by
$\chi_0(0)$ (solution at $\lambda =0$)

  \vspace{1cm}

\begin{center}

\begin{tabular}{|l|l|l|l|l|l|l|l|l|} \hline
$\lambda$&0&0.4&0.5&0.6&0.7&0.8&0.9&1 \\
\hline
$a(\lambda)$&2.338&1.99&1.896&1.801&1.704&1.604&1.502&1.398 \\
$|a'(\lambda)|$&0.84&0.9&0.92&0.94&0.96&0.98&1.01&1.04 \\
$\frac{\chi_{\lambda}(0)}{\chi_0(0)}$
&1&1.249&1.318&1.392&1.469&1.549&1.633&1.72\\ \hline
  \end{tabular}
 \end{center}

\vspace{1cm}

{\bf Table 5}\\

The factor $\rho(AF)$, as a function of
paramenter $\lambda$, (88), for two values of $\Lambda_{QCD}$
entering (88) via one--loop expression (17)
$$\rho=\frac{\chi_{AF}(0)}{\chi(0)}$$

  \vspace{1cm}

\begin{center}

\begin{tabular}{|l|l|l|l|l|l|l|} \hline
$\Lambda_{QCD}$;$\lambda$&0&0.4&0.6&0.7&0.8&1\\ \hline
90 MeV&1&0.976&0.92&0.884&0.838&0.76\\
140
MeV&1&0.984&0.95&0.911&0.877&0.79\\\hline
  \end{tabular}
 \end{center}

\vspace{1cm}

{\bf Table 6}\\

Mass correction ($-\Delta M$) in GeV, Eq. (112), due to the
rotating string, computed quasiclassically. The entries in
parenthesis are obtained from Eq. (113)

  \vspace{1cm}

\begin{center}

\begin{tabular}{|ll|l|l|l|} \hline
&n&0&1&2 \\
L&&&&\\\hline
1&&0.082&0.014&0.005 \\
&&(0.066)&&\\ \hline
2&&0.119&0.035&\\
&&(0.120)&&\\  \hline
3&&0.164 & &  \\
&&(0.167)&& \\ \hline
  \end{tabular}
 \end{center}

\newpage

{\bf Table 7}\\

Meson masses for the Hamiltonian (85)
 with the parameters
$$
\alpha_s=0.39
$$
$$\sigma=0.17 GeV^2
$$
$$
C_0=-482 MeV
$$

  \vspace{1cm}

\begin{center}

\begin{tabular}{|l|l|l|} \hline
meson& mass (theory) in MeV& mass
(experiment) in MeV\\ \hline
$\rho(1S)$&769 (fit) &$768.5\pm0.6$\\
$\rho(2S)$&1500 &$1465\pm25$\\
$\rho(3S)$&2035 &$2149\pm17$\\
$\pi(2S)$&1325 &$1300\pm100$\\
$\pi(3S)$&1895 &$1861\pm13$\\
$\bar M(1^3P_j)$&1200 &$1230\pm22=M_{cog}(1^3P_j)$\\
&&candidates$^{a)}$\\
$\bar M'2(^3P_j)$&1790
&$a_2(1700),a_1(1700)$\\
$\bar M^{''}3(^3P_j)$&2265&$a_1(2100),a_0(2050)$\\ \hline
\end{tabular}
 \end{center}
$^{a)}$ T.Barnes hep-ph/9907259

\vspace{1cm}

{\bf Table 8}\\

Compilation of lattice results for exotic $1^{-+}$ hybrid masses
 (in GeV) compared to calculations of the present approach
\vspace{1cm}

\begin{center}

\begin{tabular}{|l|l|l|l|l|} \hline
&$M(\bar b bg)-M(\bar bb)$&
$M(\bar ccg)-M(\bar cc)$&
$M(\bar ssg)$&
$M(\bar u dg)$\\
\hline
&1.14[a]&1.34[e]&2.00[g]&1.88[g]\\
&1.3[b]&1.22[f]&2.17[e]&1.97[e]\\
Lattice$^{*)}$&1.54[c]&1.323[c]&&1.90[i]\\
&1.49[d]&&&2.11[f]\\
\hline
Present&&&&\\
approach&1.5&1.45&2.34&2.16\\ \hline
\end{tabular}
 \end{center}
$^{*)}$ From D.Toussaint, hep-ph/9909088; [a],...[g] refer to
sources in the paper.

\newpage
{\bf Table 9}\\

Lattice data on glueball masses (in GeV) compared to the present
approach [60]\\
 \vspace{1cm}
 \begin{center}
 \begin{tabular}{|l|l|l|l|l|} \hline
 $J^{PC}$&[61]&[62]&[63]&Present
 approach [60]$(\sigma_f=0.23 GeV^2)$\\\hline
 $0^{++}$&$1.73\pm 0.13$&1.74&1.69&1.58 \\\hline
 &$2.67\pm0.31$&3.15&2.48&2.71 \\\hline
 $2^{++}$&$2.4\pm0.15$&2.48&&2.59 \\\hline
 &&3.22&&3.73 \\ \hline
 $0^{-+}$&$2.59\pm0.17$&2.38&2.54&2.56\\
 &$3.64\pm0.24$&&&3.77 \\\hline
 $2^{-+}$&$3.1\pm0.18$&3.38&3.31&3.03\\
&$3.89\pm0.23$&&&4.15 \\\hline
 $3^{++}$&$3.69\pm0.22$&4.31&4.27&3.58 \\\hline
 $1^{--}$&$3.85\pm0.24$&&&3.49 \\\hline
 $2^{--}$&$3.93\pm0.23$&&&3.71 \\\hline
 $3^{--}$&$4.13\pm0.29$&&&4.03 \\\hline
 $1^{+-}$&$2.94\pm0.17$&3.03&& \\\hline
 $2^{+-}$&$4.14\pm0.25$&&4& \\\hline
 $0^{+-}$&$4.74\pm0.3$&&& \\\hline
 \end{tabular}
 \end{center}

\vspace{1.5cm}

{\bf Table 10}\\

Characteristics of heavy--light mesons in comparison with
experiment and lattice data
  \vspace{1cm}

\begin{center}

\begin{tabular}{|l|l|l|l|l|l|l|l|} \hline
meson&$B$& $B^*$&$B_s$&$D$&$D^*$&$D_s$&$D^*_s$\\\hline
$M-m_Q$&0.479&0.539&0.535&0.496&0.630&0.532&0.666\\  \hline
$M$&5.279&5.339&5.335&1.896&2.030&1.932&2.066\\\hline
$m_{exp}$&5.279&5.324&5369&1.869&2.010&1.968&2.112\\\hline
$f_M(Gev)$;&0.183&&&0.221&&0.264&\\
$\sqrt{\sigma}=0.427$&&&&&&&\\\hline
$f_M^{(lat)}$&0.210&&0.251&&&&\\
$\sqrt{\sigma}=0.427$&&&&&&&\\ \hline
\end{tabular}
 \end{center}

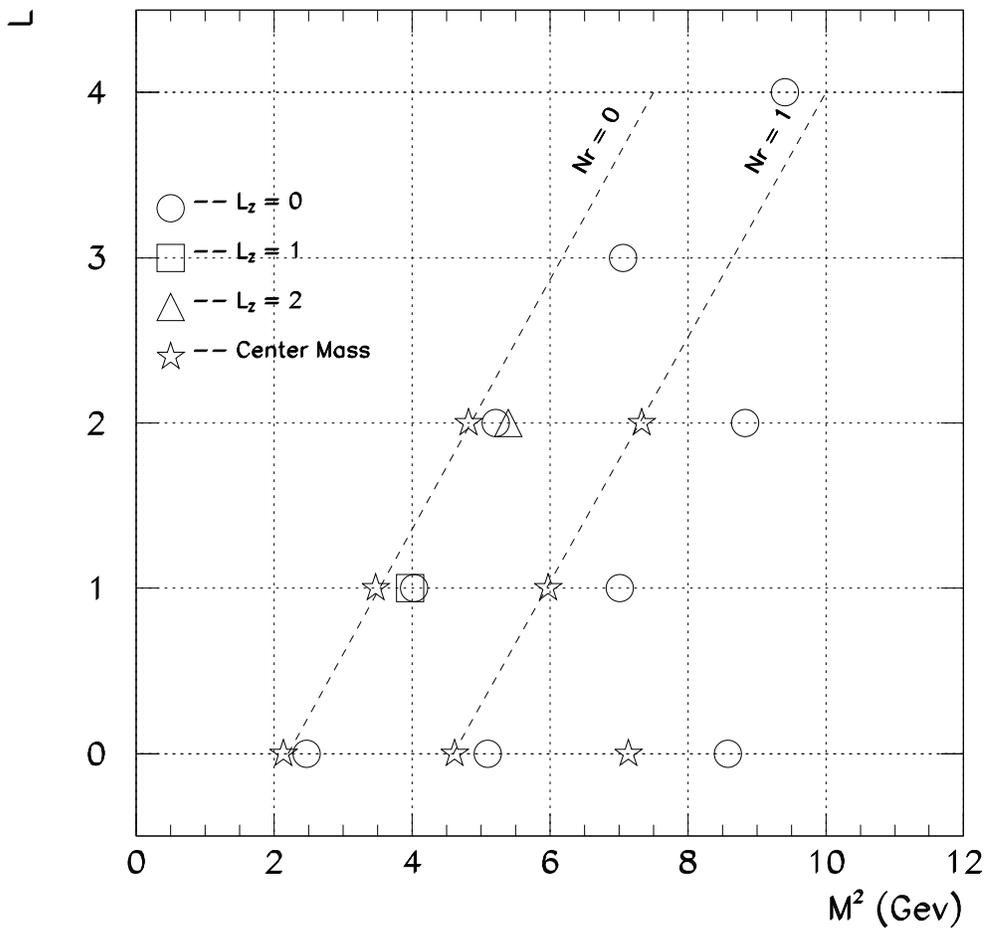
\begin{figure}[t]
\caption{Quasiclassical spectrum of Hamiltonian  (67) for
$m=0$ and $\sigma=0.2 GeV^2$. The leading experimental Regge
trajectory in  angular momentum $L$ is given in the upper plot
for comparison. Theoretical prediction for the $\rho$--meson
mass, with color Coulomb and spin  effects included is shown
by the dot at $L=0$ and does not violate the straight--line
behaviour of the leading trajectory}
\label{fig1}
\end{figure}

\begin{figure}
\caption{
The Chew-Frautschi plot with masses computed via the light--cone
(circles, squares and triangle for $L_{z} = 0,1,2$ respectively) and
the c.m. Hamiltonian (stars). The systematic overall mass shift is
seen as a divergence of straight lines passing through circles and stars.
The states with high $L$ or high $N_r$ (daughter trajectories) are
numerically less reliable and not shown. }
\label{fig2}
\end{figure}

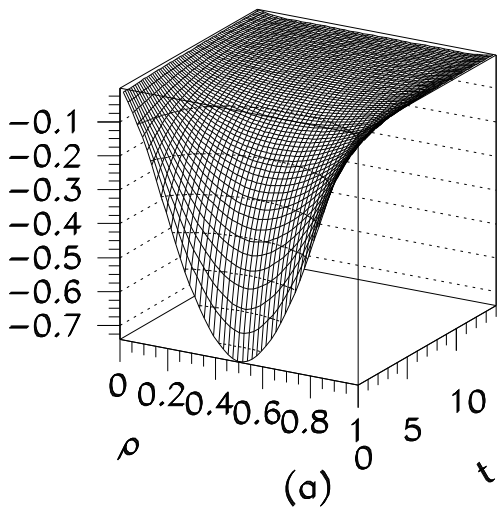
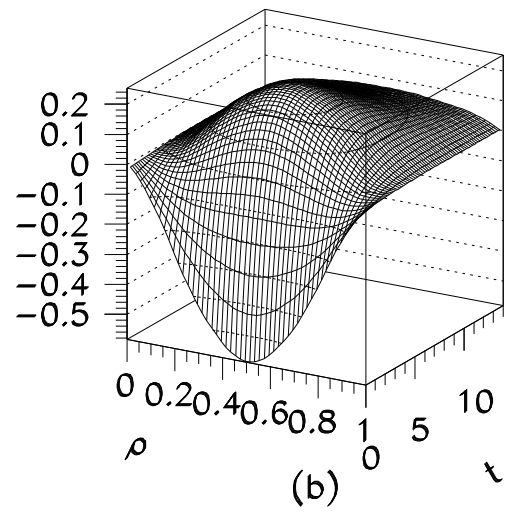
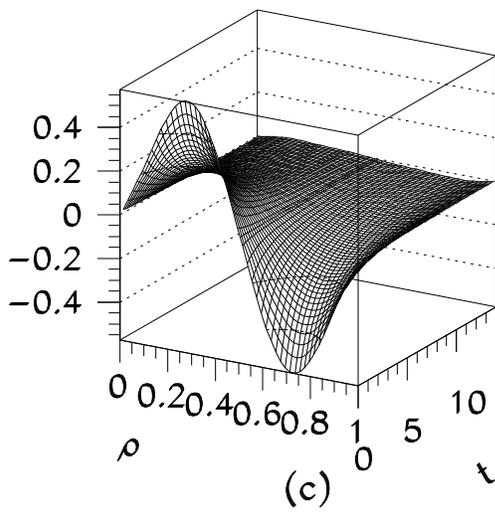
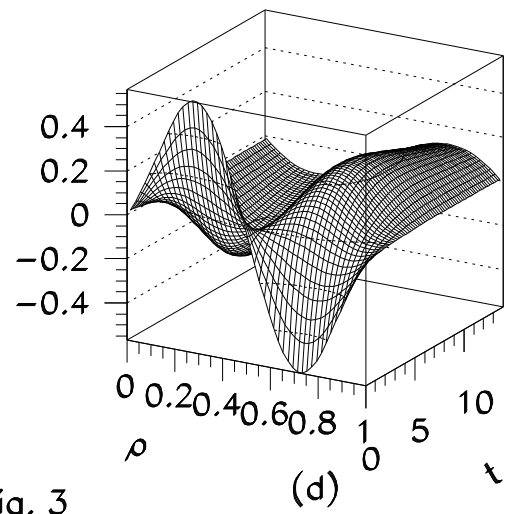
\begin{figure}
\caption{
The 3d plots of wave functions of the four lowest states of light--cone
Hamiltonian for zero quark masses. Coordinates on horizontal plane are
$0 \leq \rho \leq 1$ , $0 \leq t \leq 15$ (in units of $\sigma^{-1}$). }
\label{fig3}
\end{figure}

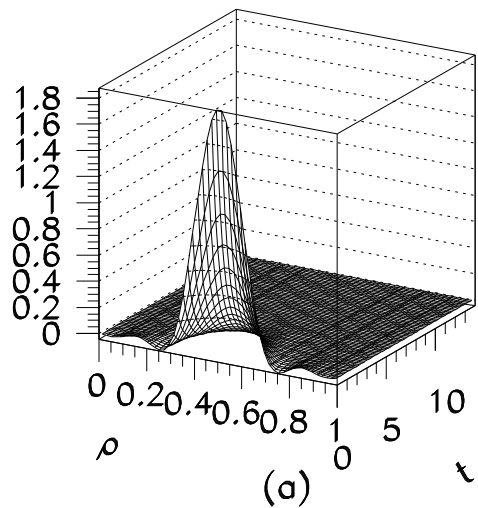
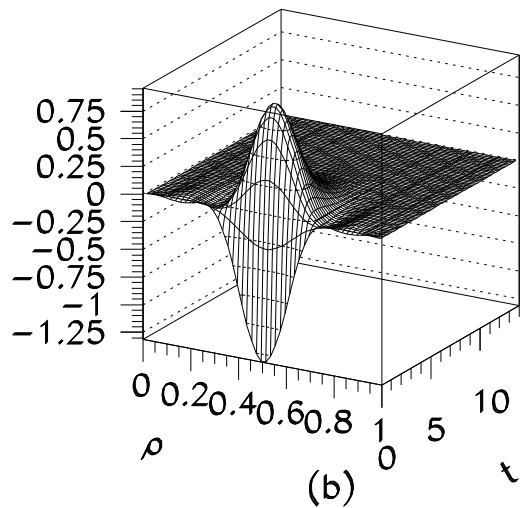
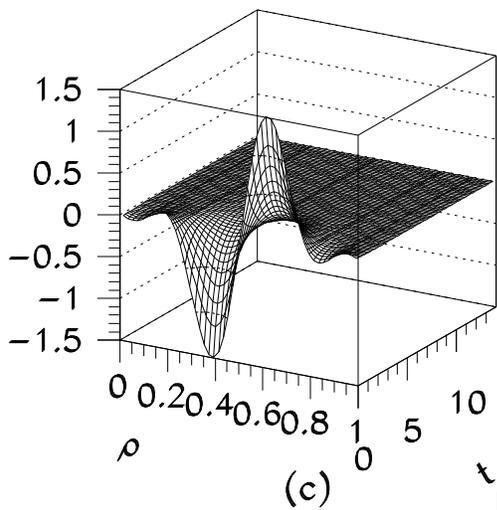
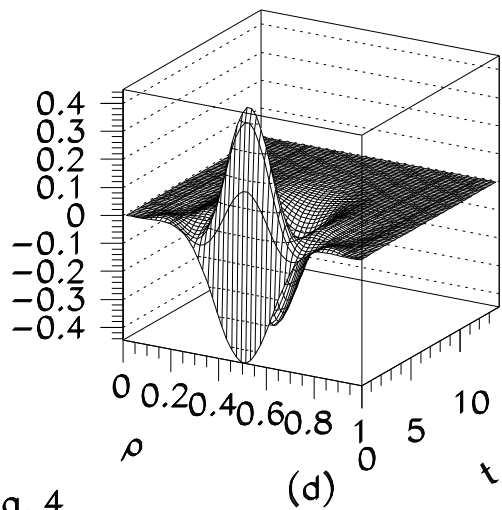
\begin{figure}
\caption{
The same as in Fig.~3 but for heavy quark masses, $m_1 = m_1 = 5$ GeV. }
\label{fig4}
\end{figure}

\begin{figure}
\caption{
The 3d plots of the ground--state wave functions $\Psi(\rho , t)$,
computed via the light--cone Hamiltonian (upper part) and via the c.m.
Hamiltonian, with the standard substitution (\ref{g}) (lower part) for
zero quark masses. }
\label{fig5}
\end{figure}


\begin{figure}
\caption{
The quark--distribution function $q(\rho)$ computed with light--cone
wave--functions for the cases considered. }
\label{fig6}
\end{figure}

 
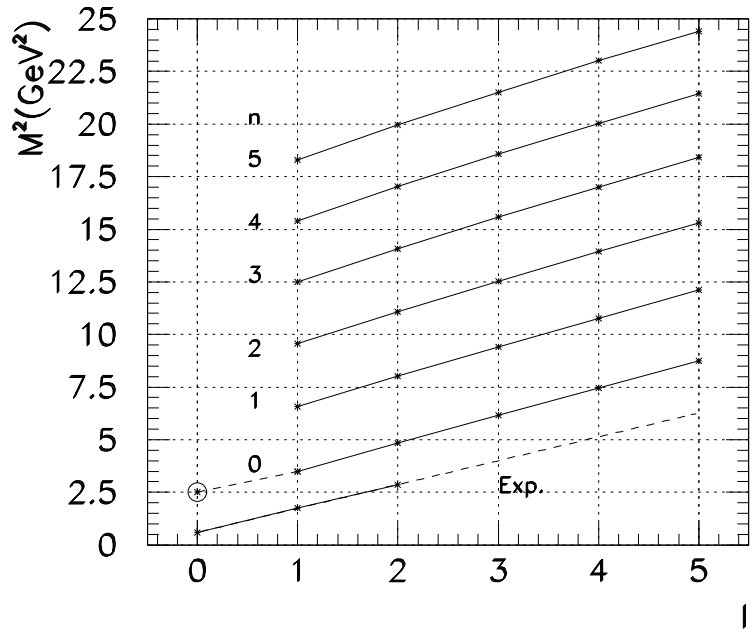
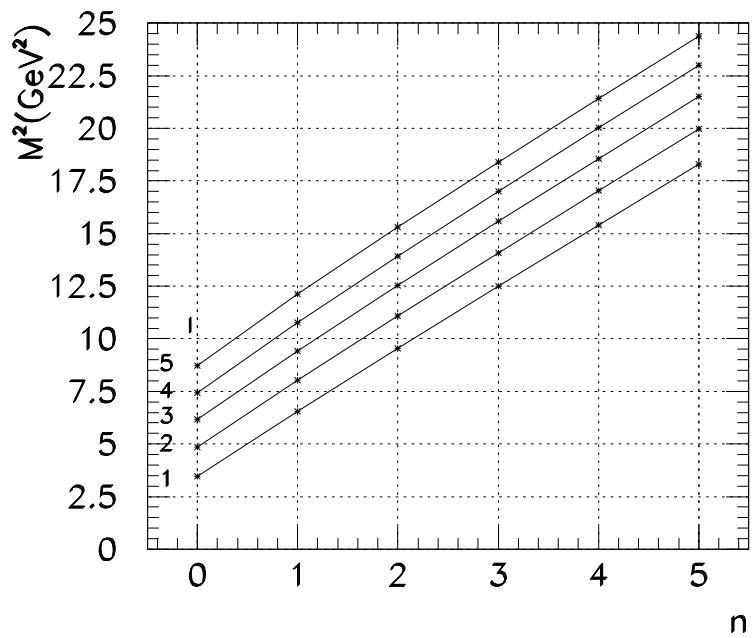
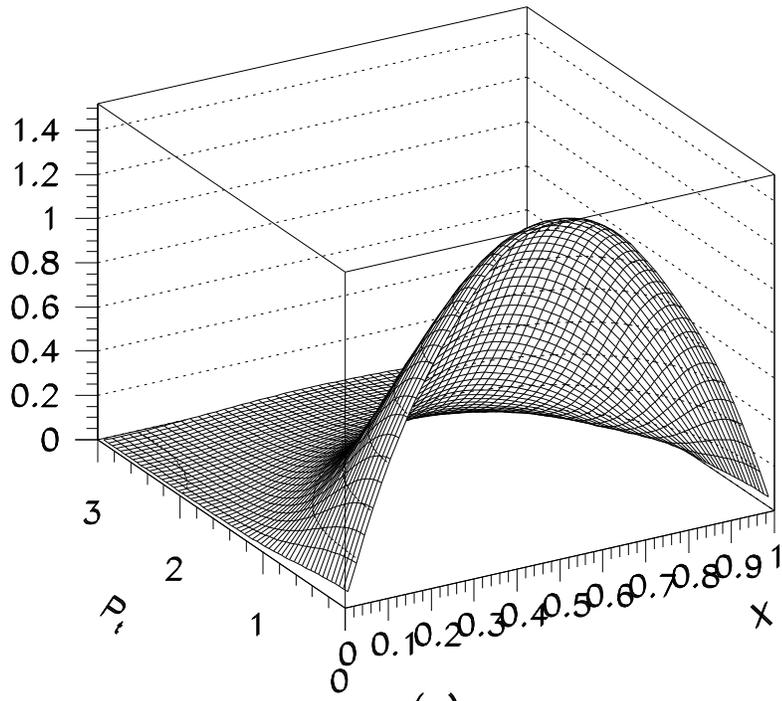
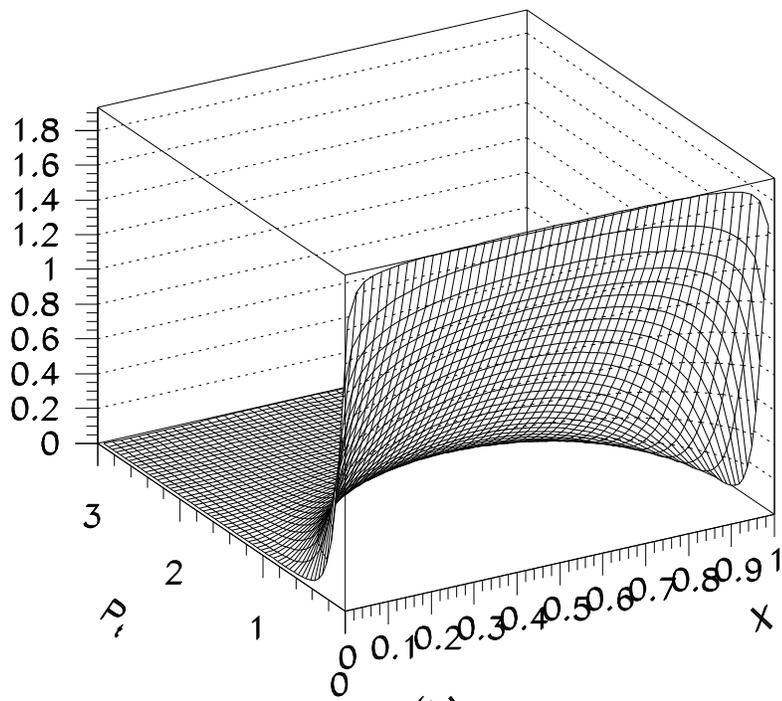
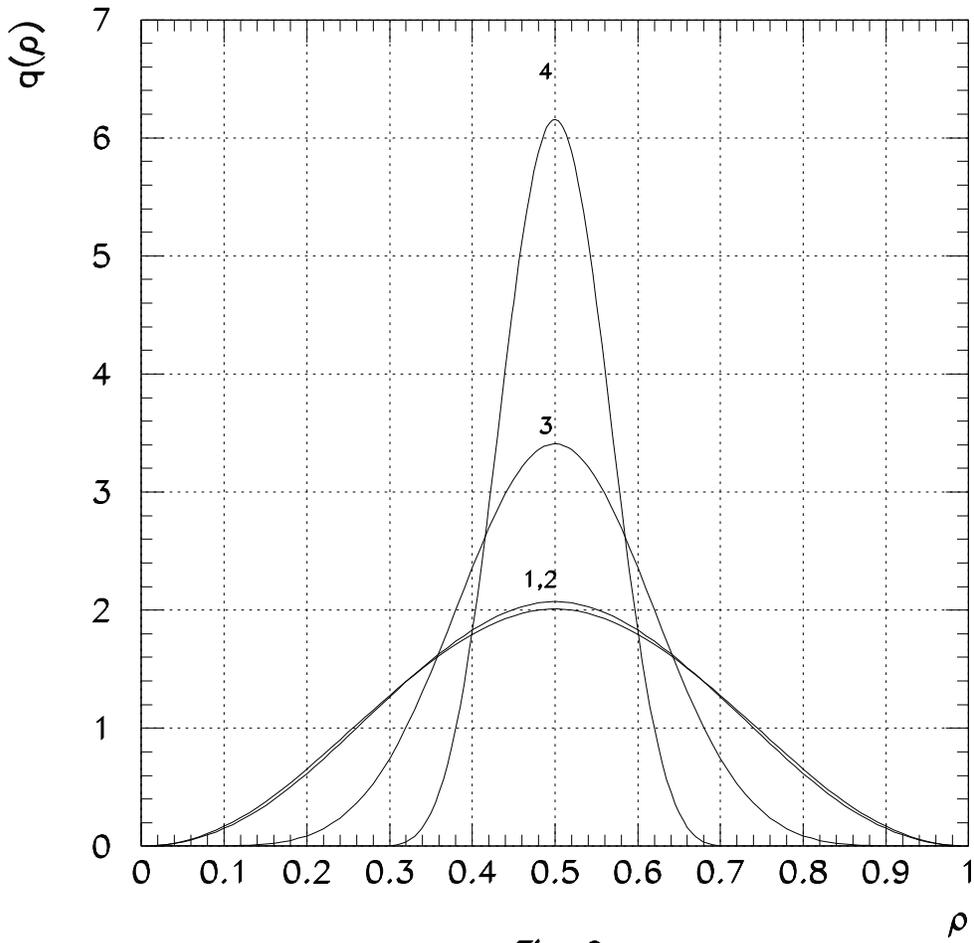
\end{document}